\documentclass[conference]{IEEEtran}
\IEEEoverridecommandlockouts
\usepackage{cite}
\usepackage{amsmath,amssymb,amsfonts}
\usepackage{algorithmic}
\usepackage{graphicx}
\usepackage{textcomp}
\usepackage{xcolor}
\def\BibTeX{{\rm B\kern-.05em{\sc i\kern-.025em b}\kern-.08em
    T\kern-.1667em\lower.7ex\hbox{E}\kern-.125emX}}
    
\usepackage{multirow}
\usepackage{gensymb}
\usepackage{threeparttable}
\usepackage[caption=false]{subfig}
\usepackage{xcolor}
\newcommand{\ineq}[1]{\footnotesize$#1$\normalsize}{}
\newcommand{\mr}[1]{\textcolor{black}{#1}}

\definecolor{mygrey}{gray}{0.80}

\newcommand{\ipc}{\text{{interprocessor communication}}}{}
\newcommand{\pb}{\text{{SentryOS}}}{}
\newcommand{\base}{\text{{Baseline}}}{}
{}
{}
{}
\newcommand{\nsoc}{\text{{NSoC}}}{}
\newcommand{\grai}{\text{{GrAI}}}{}
\newcommand{\tech}{\text{{PRISM}}}{}
\newcommand{\dpar}{\text{{operation parallelism}}}{}

\usepackage[caption=false]{subfig}
\usepackage{multicol}
\newtheorem{Tlemma}{Lemma}

\newtheorem{Tdef}{Definition}
\newenvironment{Definition}[1]
    {\begin{Tdef}\noindent\textsc{(#1)}\itshape}
    {\end{Tdef}}
\newcommand{\actor}[1]{\textbf{\emph{{#1}}}}
\usepackage{multirow}
\DeclareMathAlphabet{\mymathbb}{U}{BOONDOX-ds}{m}{n}
\usepackage{amsmath}
\usepackage{pifont}
\let\oldding\ding
\renewcommand{\ding}[2][1]{\scalebox{#1}{\oldding{#2}}}

\usepackage{amsfonts}
\usepackage[linesnumbered,ruled,noend]{algorithm2e}
\SetKwRepeat{Do}{do}{while}
\DontPrintSemicolon
\SetKwProg{Def}{def}{:}{}
\SetKwProg{Class}{class}{:}{}
\usepackage{xcolor}

\SetCommentSty{mycommfont}    
\usepackage{pifont}
\usepackage{url}
\usepackage{comment}

\begin{document}
\bstctlcite{IEEEexample:BSTcontrol}

\title{Real-Time Scheduling of Machine Learning Operations on Heterogeneous Neuromorphic SoC}

\author{\IEEEauthorblockN{Anup Das}
\IEEEauthorblockA{\textit{Electrical and Computer Engineering} \\
\textit{Drexel University}\\
Philadelphia, USA \\
anup.das@drexel.edu}
}

\maketitle

\begin{abstract}
Neuromorphic Systems-on-Chip (\nsoc{s}) are becoming heterogeneous by integrating general-purpose processors (GPPs) and neural processing units (NPUs) on the same SoC.
For embedded systems, an \nsoc{} may need to execute
user applications built using a variety of machine learning models.
We propose a real-time scheduler, called \tech{}, which can schedule machine learning models on a heterogeneous \nsoc{} either individually or concurrently to improve their system performance.
\tech{} consists of the following four key steps.
First, it constructs an \ipc{ (IPC)} graph of a machine learning model from a mapping and a self-timed schedule.
Second, it creates a transaction order for the communication actors and embeds this order into the IPC graph.
Third, it schedules the graph 
on an \nsoc{} by overlapping communication with the computation.
Finally, it uses a Hill Climbing heuristic to explore the design space of 
mapping 
operations on GPPs and NPUs to improve the performance.
Unlike existing schedulers which 
use only the NPUs of an \nsoc{}, 
\tech{} improves performance by enabling batch, pipeline, and \dpar{} via exploiting a platform's heterogeneity.
For use-cases with concurrent applications, \tech{} uses 
a heuristic resource sharing strategy and a non-preemptive scheduling
to reduce the expected wait time before concurrent operations can be scheduled on contending resources.
Our extensive evaluations with 20 machine learning 
workloads show that \tech{} significantly improves the performance per watt for both individual applications 
and 
use-cases when compared to state-of-the-art schedulers. 
\end{abstract}

\begin{IEEEkeywords}
neuromorphic system-on-chip (NSoC), spiking deep convolution neural network (SDCNN), interprocessor communication (IPC), hill climbing, self-timed execution.
\end{IEEEkeywords}

\section{Introduction}\label{sec:introduction}
\IEEEPARstart{E}{vent}-driven neuromorphic devices are 
a promising solution to accelerate Spiking Deep Convolutional Neural Networks (SDCNNs)~\cite{han2016energy,sengupta2019going,cao2015spiking}.
These are a type of Spiking Neural Networks (SNNs) that replicate the neural architecture of a conventional CNN with two major differences~\cite{maass1997networks}. 
First, layers of an SDCNN communicate by exchanging spikes instead of tensors. 
Second, each layer implements the leaky integrate-and-fire function instead of a sigmoidal function (Tanh or ReLU).
Due to these differences, converting a CNN to an equivalent SDCNN is not straightforward, and it requires layer-specific optimizations~\cite{datta2021can,xing2019homeostasis,fang2021deep,tecs2021,mdpi2022,jolpe2018}. 

Our \underline{objective} is to schedule SDCNNs on a neuromorphic device that
consists of neural processing units (NPUs) to accelerate operations such as matrix multiplication, average/max pooling, batch normalization, layer flattening, residual computation, and concatenation~\cite{mead1990neuromorphic,li2021escalate,shao2021memory,mrazek2019alwann,springer2018}.
These devices are now integrated with general purpose processors (GPPs) and other hardware subsystems on the same system-on-chip (SoC) to implement a complete system.
Neuromorphic systems-on-chip (\nsoc{s}) are becoming the key enabler for embedded systems 
to accelerate machine learning-based user applications by scheduling their operations on NPUs. Scheduling consists of three steps --- mapping, which involves assigning operations to different NPUs, ordering, which involves determining the execution order of operations and their communication, and timing, which involves specifying the precise start time of an operation on an NPU~\cite{SB00,glsvlsi2018,isvlsi2019,lctes2020,iccad2021}.

Table~\ref{tab:nsocs} reports \underline{event-driven} \nsoc{} platforms (top four rows) and their system software to schedule SDCNN operations.\footnote{There are also tensor-based \nsoc{s} such as Qualcomm's Snapdragon~\cite{snpe} and Nvidia's Jetson AGX Xavier~\cite{agx}. These \nsoc{s} are evaluated in~\cite{ignatov2018ai}.} It also reports a few standalone neuromorphic devices and their software (bottom four rows). These devices are not currently part of any SoC but reported here for completeness.

\vspace{-10pt}
\begin{table}[h!]
	\renewcommand{\arraystretch}{1.0}
	\setlength{\tabcolsep}{10pt}
	\caption{Recently introduced \nsoc{} platforms.}
	\label{tab:nsocs}
	\vspace{-5pt}
	\centering
	\begin{threeparttable}
	{\fontsize{6}{10}\selectfont
		\begin{tabular}{c|cc|c}
			\hline
			\textbf{\nsoc{}} & \textbf{\# NPUs} & \textbf{GPPs} & \textbf{Software} \\
			\hline
			Loihi~\cite{loihi} & 128 & 3 x86 CPUs & LAVA~\cite{loihi_mapping}\\
			Akida~\cite{akida} & 80 & 1 x86 CPU & MetaTF~\cite{metatf}\\
			GrAI~\cite{grai} & 144 & 2 ARM CPUs & GrAIFlow~\cite{grai}\\
			SPECK~\cite{speck} & $>$ 16 & 1 GPU & CTXCTL~\cite{speck}\\
			\hline
			SpiNNaker~\cite{spinnaker} & 144 & -- & PACMAN~\cite{pacman} \\
			$\mu$Brain~\cite{date2022} & 64 & -- & SentryOS~\cite{date2022}\\
			Tianji~\cite{tianji} & 6 & -- & NEUTRAMS~\cite{neutrams} \\
			DYNAPs~\cite{dynapse} & 4 & -- & SpiNeMap~\cite{date2018,tvlsi2019,esl2020}\\
			\hline
	\end{tabular}}
	\end{threeparttable}
\end{table}
\vspace{-5pt}

\begin{figure}[h!]
	\centering
	\vspace{-10pt}
	\centerline{\includegraphics[width=0.99\columnwidth]{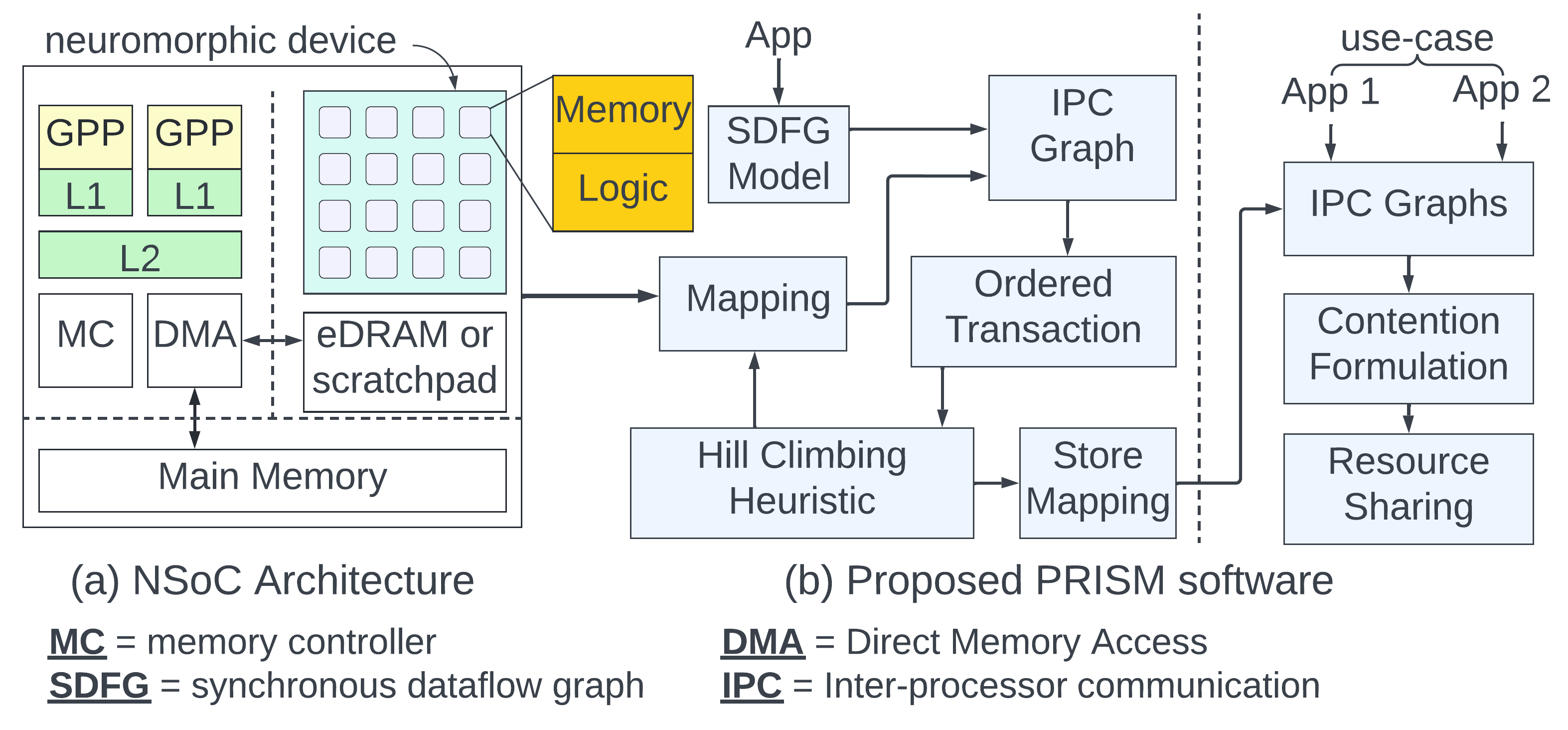}}
	\vspace{-10pt}
	\caption{(a) An \nsoc{} platform with GPPs and NPUs. (b) Proposed scheduler \tech{} which schedules both individual applications and use-cases.}
	\vspace{-5pt}
	\label{fig:nsoc}
\end{figure}

Figure~\ref{fig:nsoc}a shows the architecture of an \nsoc{} with GPPs and NPUs, where 
NPUs are arranged in an XY mesh.
Internally, each NPU consists of 
circuitries to perform neural computations and
memory 
to store synaptic weights. 
Unlike a GPP, an NPU does not require frequent memory accesses due to its self-contained memory architecture and therefore, it does not suffer from the memory bandwidth bottleneck.
Due to its small hardware area, an \nsoc{} may integrate several NPUs (see Table~\ref{tab:nsocs}) to process a batch of data at once. This is called {batch parallelism}. A system software exploits this parallelism to improve the throughput.
Following are the three key limitations of existing schedulers that motivate this work.

First, existing schedulers do not exploit all forms of parallelism that exist in an \nsoc{}.
For instance, an \nsoc{} can support {pipeline parallelism}, where the processing of operations 
of subsequent input data can be 
overlapped in time by utilizing the platform's heterogeneous computing resources.
It can also support {\dpar{}}, where independent operations acting on the same input data can be scheduled concurrently on multiple computing resources.
Figure~\ref{fig:motivation} illustrates the difference between these three forms of parallelism.


Second,
machine learning is an evolving area of research and new models are introduced to the community on a regular basis.
We analyze five models -- SqueezeNet~\cite{squeezenet}, InceptionV3~\cite{inception}, U-Net~\cite{unet}, BERT~\cite{bert}, and ConvLSTM~\cite{convlstm}, for their support on existing \nsoc{s}.
Table~\ref{tab:supported_model_summary} summarizes these results. While all \nsoc{s} support SqueezeNet, none of them support U-Net, BERT, and ConvLSTM.
On the other hand, \grai{} is the only hardware that supports InceptionV3.
Existing schedulers cannot map a machine learning model on an \nsoc{} if \textit{any} of its operations is not supported on the \nsoc{'s} NPU.

\vspace{-10pt}
\begin{table}[h!]
	\renewcommand{\arraystretch}{1.0}
	\setlength{\tabcolsep}{2pt}
	\caption{Analysis of existing \nsoc{} platforms.}
	\label{tab:supported_model_summary}
	\vspace{-5pt}
	\centering
	\begin{threeparttable}
	{\fontsize{6}{10}\selectfont
		\begin{tabular}{c|ccccc|c}
			\hline
			\textbf{\nsoc{}} & \textbf{SqueezeNet} & \textbf{InceptionV3} & \textbf{U-Net} & \textbf{BERT} & \textbf{ConvLSTM} & \textbf{Custom Model}\\
			\hline
			Loihi & $\surd$ & $\times$ & $\times$ & $\times$ & $\times$ & $\times$\\
			Akida & $\surd$ & $\times$ & $\times$ & $\times$ & $\times$ & $\times$\\
			GrAI & $\surd$ & $\surd$ & $\times$ & $\times$ & $\times$ & $\times$\\
			SPECK & $\surd$ & $\times$ & $\times$ & $\times$ & $\times$ & $\times$\\
			\hline
	\end{tabular}}
	\end{threeparttable}
\end{table}
\vspace{-5pt}

Third, an \nsoc{} may need to execute different machine learning applications concurrently to satisfy user demands.
Imagine a use-case of running social media services (image classification using MobileNet or InceptionV3) while listening to music (audio processing using BERT or ConvLSTM) on a cell phone. Existing schedulers cannot schedule more than one machine learning application at the same time.

We propose \tech{}, a \underline{p}erformance-oriented \underline{r}eal-t\underline{i}me \underline{s}cheduler for \underline{m}achine learning operations on a heterogeneous \nsoc{}.
Following are our key \textbf{contributions}.

\begin{enumerate}
    \item We propose a mechanism to construct an \ipc{} graph from a machine learning model using a mapping and a self-timed schedule of its operations on the computing resources. We exploit the expressiveness of a synchronous dataflow graph (SDFG) to represent an IPC graph.
    \item We propose a transaction partial order algorithm to create a transaction order for the inter-processor communications of an IPC graph. We embed this transaction order into the graph and schedule it on an \nsoc{} so that communication is overlapped with the computation.
    \item We propose a Hill Climbing heuristic to explore the design space of scheduling
    machine learning operations on 
    the computing resources and create opportunities for batch, pipeline, and \dpar{} via exploiting the platform's heterogeneity.
    \item We propose a probabilistic formulation to estimate the performance slowdown due to resource contention. 
    We incorporate this formulation within a use-case mapping framework that uses a heuristic resource sharing strategy and a non-preemptive scheduling to map applications of a use-case on an \nsoc{}. \tech{} minimizes the performance slowdown by reducing the expected wait time for operations scheduled on contending resources.
\end{enumerate}

Our extensive evaluations with 20 machine learning 
workloads and five use-cases show that \tech{} significantly improves the performance and performance per watt for both individual applications 
and multi-application use-cases 
when compared to state-of-the-art schedulers. 

To the best of our knowledge, \tech{} is the only scheduler that can schedule 
machine learning applications (standard or custom)
either individually or concurrently by exploiting the heterogeneity of an \nsoc{} platform.

\textbf{Why Spiking?}
\tech{} primarily deals with trained spiking CNN (i.e., SDCNN inference) models because spiking hardware platforms are energy-efficient due to their event-driven operations.
Therefore, these platforms are ideal for embedded systems and other environments where machine learning tasks may need to be performed within an energy budget.
Nevertheless, with simple modifications \tech{} can also schedule operations of a conventional CNN.

\textbf{Overview of \tech{}:}
Figure~\ref{fig:nsoc}b shows an overview of \tech{}. It has two main components -- individual application exploration (left) and use-case mapping strategy (right). For mapping individual applications, \tech{} records all Pareto-optimal schedules generated from the proposed Hill Climbing heuristic. 
It selects a schedule that gives the highest performance for a given energy constraint.
For mapping use-cases,
\tech{} finds the best strategy to share resources amongst the concurrent applications. \tech{} improves the quality of experience by minimizing the time from when an application is invoked to when it starts executing.


\section{Background and Motivation}\label{sec:background}
\begin{figure}[h!]
	\centering
	\vspace{-10pt}
	\centerline{\includegraphics[width=1.00\columnwidth]{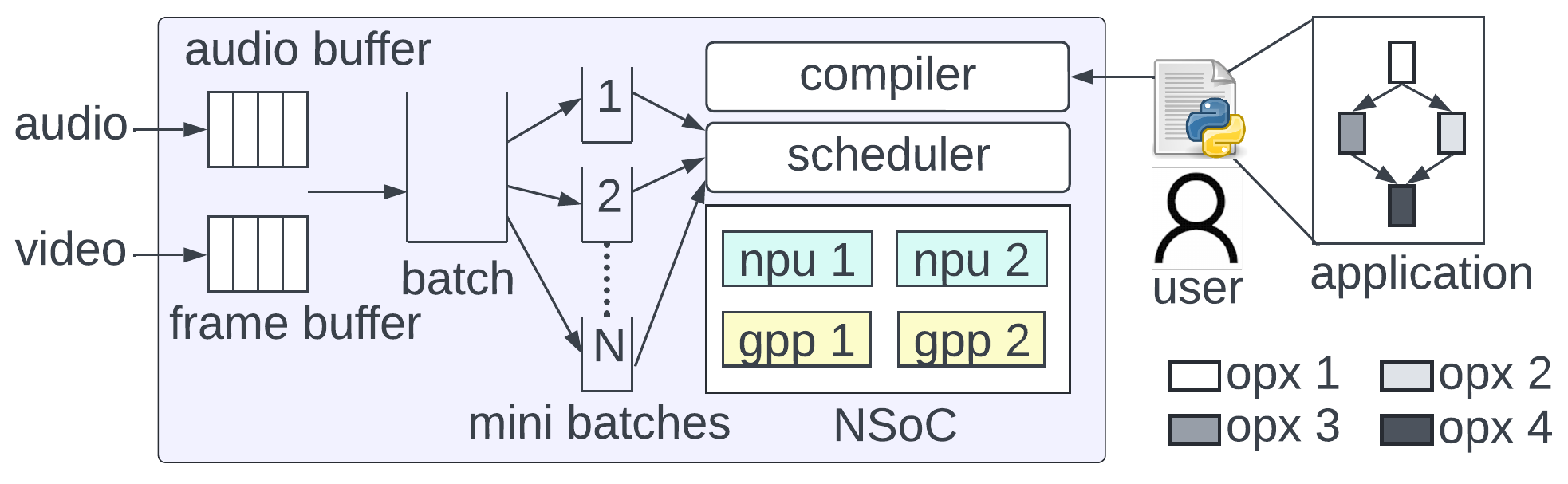}}
	\vspace{-10pt}
	\caption{An embedded platform running a machine learning application.}
	\vspace{-5pt}
	\label{fig:platform}
\end{figure}

To illustrate 
the different forms of parallelism in an \nsoc{}, 
Figure~\ref{fig:platform} shows an embedded platform performing audio and video processing.
A compiler compiles a user application into an intermediate representation, which
is shown to the right with four operations (opx 1-4).
A scheduler schedules these operations on the computing resources of the hardware.
On the input front,
audio/video frames are first streamed into buffers.
Once a batch is ready, the scheduler distributes it into mini-batches and forwards them to the computing resources, one at time. The number of frames in each mini-batch is equal to the number of NPUs so they can be scheduled in parallel.

\begin{figure}[h!]
	\centering
	\vspace{-10pt}
	\centerline{\includegraphics[width=0.99\columnwidth]{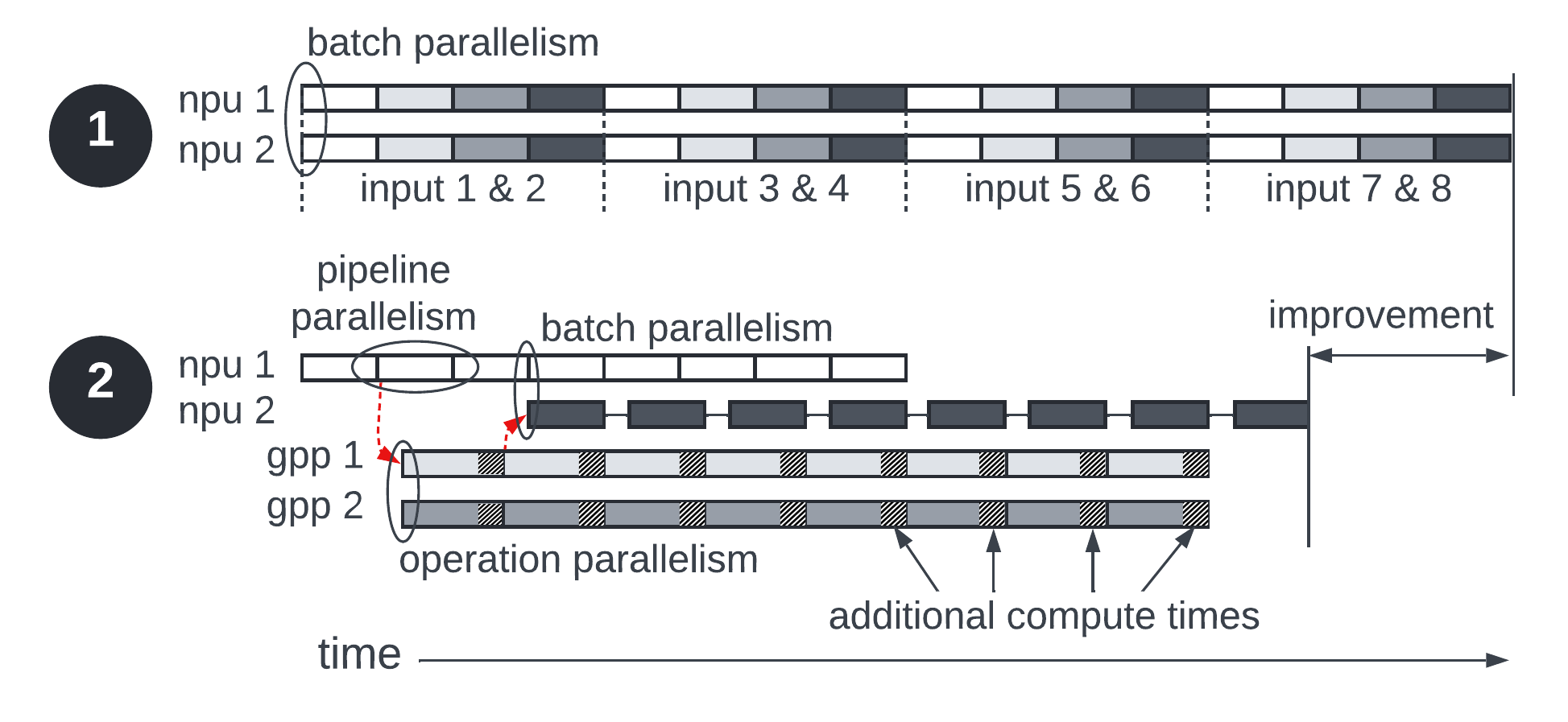}}
	\vspace{-10pt}
	\caption{An example showing the different forms of parallelism in an \nsoc{}.}
	\vspace{-5pt}
	\label{fig:motivation}
\end{figure}

Figure~\ref{fig:motivation} illustrates two different schedules. 
In \ding{182}, we show 
an existing scheduler 
that uses two NPUs.
Each NPU operates
on a single input data from a mini-batch and performs all four operations shown in Figure~\ref{fig:platform} (right). Once done, these NPUs operate on the next mini-batch. 
This is \underline{batch parallelism}. Most schedulers exploit this parallelism 
to improve the performance.

In \ding{184}, we illustrate 
how \tech{} schedules the mini-bathes.
We make the following five observations.
{First}, \tech{} uses all computing resources (two GPPs and two NPUs).
{Second}, unlike the baseline schedule where each NPU processes all four operations sequentially, here each computing resource processes a specific operation for every input data.
Essentially, \tech{} creates a four-stage pipeline with NPU~1, GPP 1, GPP 2, and NPU 2, respectively.
The processing of a new input can begin when the first pipeline stage (NPU-1) completes its execution. 
This is \underline{pipeline parallelism}.

Third, \tech{} schedules operations 2 \& 3 on the two GPPs. The figure illustrates additional computation times necessary to process these operations on a GPP compared to an NPU.
{Fourth}, operations 2 \& 3 can be scheduled in parallel as they are independent (see Figure~\ref{fig:platform}). \tech{} schedules these operations on the two GPPs. This is the \underline{operation parallelism}, which also contributes to performance improvement.

{Finally,} data exchanges are necessary between the computing resources.
To move data from a GPP to an NPU, \tech{} first copies the data from the GPP's cache to the main memory if it is not already there. Next, it initiates a data transfer from the main memory to the memory of a neuromorphic device using the DMA module. Once loaded, it distributes this data to the target NPU using the interconnect.
Typically, a neuromorphic device integrates an embedded DRAM (eDRAM) or a scratchpad memory for data storage as shown in Figure~\ref{fig:nsoc}a.
Conversely, to move data from an NPU to a GPP, \tech{} first copies the data to the main memory using the DMA module. Thereafter, a GPP loads this data into its cache using its cache placement/replacement policies.
These communications are indicated using red arrows.
\tech{} overlaps data communication with the computation, which further improves the performance. We report 2.2x higher performance compared to state-of-the-art schedulers (Section~\ref{sec:results}).


\subsection{Synchronous Dataflow Graphs}\label{sec:sdfg_intro}
Synchronous Data Flow Graphs (SDFGs, see~\cite{sdfg}) are used to model streaming applications that are implemented on a multi-processor SoC (MPSoCs, see~\cite{mpsoc}).
These graphs are used to analyze a system in terms of 
execution time~\cite{rosvall2014constraint}, throughput~\cite{ghamarian2006throughput},
buffer requirements~\cite{Stuijk06dac}, energy~\cite{tecs2014}, power~\cite{dac2013}, temperature~\cite{date2014}, and reliability~\cite{tpds2015}.

\tech{} models a machine learning application as an SDFG.
Nodes of an SDFG are called \textit{actors} and they compute by reading \textit{tokens}.
A token is the smallest unit of data communicated between actors. 
To model an SDCNN as an SDFG, we consider the SDCNN application to be composed of a set of operations (convolution, pooling, batch normalization, etc).
We represent each operation as an actor and spikes generated from these operations as tokens.
Before an actor starts its execution, it reads tokens from its input ports.
After completing its execution, it produces tokens on all of its output ports.
The number of tokens produced or consumed in one execution of an actor is called the \textit{port rate}.
Port rates are visualized as annotations on edges.
Actor execution is also called \textit{firing}, and it requires a fixed amount of time to execute on a computing unit. Edges in the graph are called \textit{channels} and they represent dependencies among actors.

An actor is called {\em ready} when it has sufficient input tokens on all its input channels and sufficient buffer space on all its output channels; an actor can only fire when it is ready.

\begin{Definition}{Actor}
{An actor $\actor{a}_i$ is the tuple $\langle I_i,O_i,\tau_i,\mu_i\rangle$ consisting of a set $I_i$ ($\subseteq Ports$) of input ports, a set $O_i$ ($\subseteq Ports$) of output ports, $\tau_i$ is the execution time of $\actor{a}_i$ and $\mu_i$ is its state space, which is the memory required to store synaptic weights, and input and output spikes.}
\end{Definition}

The execution time \ineq{\tau_i} of actor $\actor{a}_i$ is the tuple with its execution time on different computing resources.

\begin{Definition}{SDFG}
{An SDFG is a directed graph $G = (A,C)$ consisting of a finite set $A$ of actors and a finite set $C\subseteq Ports^2$ of channels. The source of channel $ch_i^j \in C$ is an output port of actor $\actor{a}_i$, the destination is an input port of actor $\actor{a}_j$. All ports of all actors are connected to precisely one channel.
Channels connected to input and output ports of an actor $\actor{a}_i$ are denoted by $InC(\actor{a}_i)$ and $OutC(\actor{a}_i$), respectively.
}
\end{Definition}

\begin{figure*}[h!]
	\centering
	\centerline{\includegraphics[width=1.99\columnwidth]{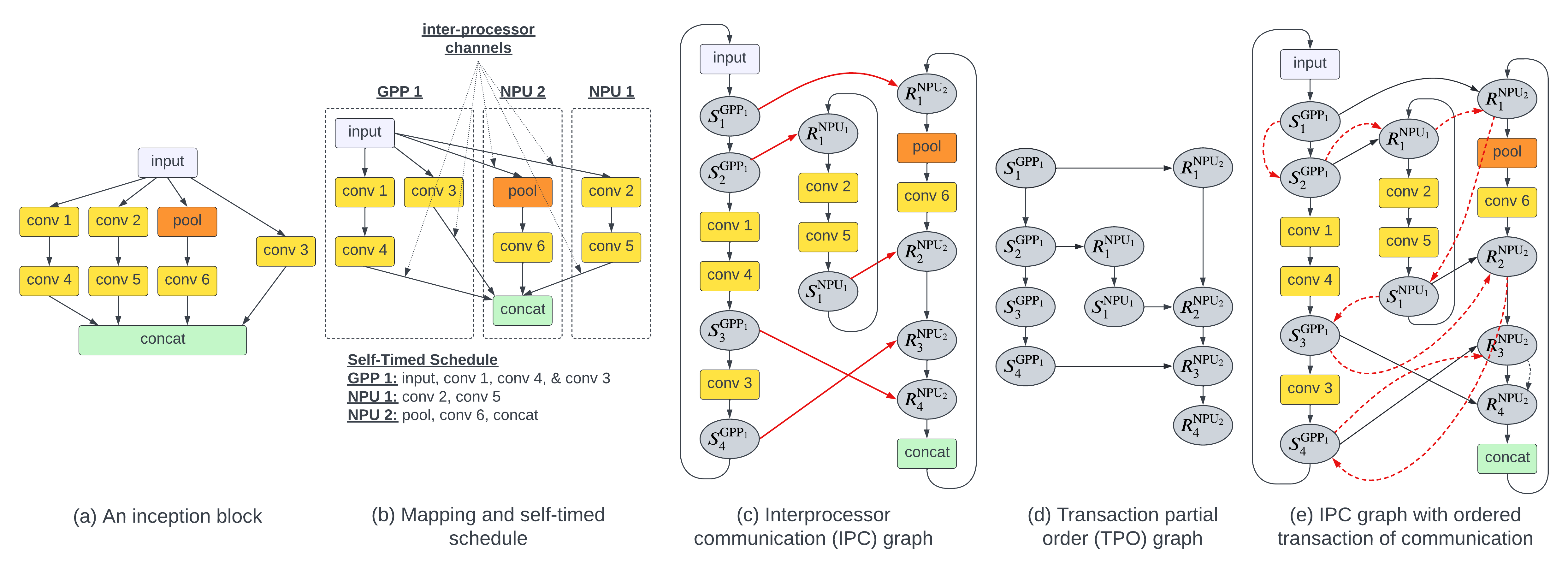}}
	\caption{(a) An inception block. (b) Self-timed schedule of the SDFG representing the inception block. (c) \ipc{ (IPC)} graph representing the self-timed execution. (d) A transaction partial order (TPO) graph formed by eliminating the computing actors. This graph is used to create a transaction order of these communication actors. (e) The IPC graph embedded with the transaction order.}
	\label{fig:ipc}
\end{figure*}

In our formulation, channels are delayless, i.e., tokens produced in one invocation of an actor are consumed within the iteration.
One important performance property of an SDFG is its \textit{throughput}, which is defined as the inverse of its long-term period. A period is the average time needed for one iteration of an SDFG. An iteration is defined as the minimum non-zero execution such that the original state of an SDFG is obtained. 
This is our performance metric.

\subsection{Performance Estimation of an SDFG}\label{sec:performance}
The long-term throughput of an SDFG can be computed by analyzing its maximum cycle mean (MCM). We introduce the following definitions.

\begin{Definition}{{Digraph}}
	The digraph $\Gamma(T)$ of a $n\times n$ matrix $T$ with entries defined in $\mathbb{R}_{\text{max}}$ is the tuple $\langle A,E\rangle$, where $A$ is the set of vertices, i.e., $A = \{1,2,\cdots n\}$ and $E$ is the set of connected ordered arcs between vertices i.e., $E = \{(i,j)~|~T_{i,j}\neq -\infty\}$.
\end{Definition}

\begin{Definition}{{Walk}}
	A walk $w$ in digraph $\Gamma(T)$ is the sequence of arcs $(x_1,x_2)(x_2,x_3)\cdots(x_{k-1},x_k)$;  head of an arc in the sequence is either the start vertex of the walk or tail vertex of a preceding arc; and the tail vertex of an arc in the sequence is either the end vertex of the walk or head vertex of a succeeding arc. Weight of the walk is given by
	\begin{equation}
	\label{eq:weight}
	\footnotesize|w|_T =  T_{x_1 x_2} + \cdots T_{x_{k-1} x_k}
	\end{equation}
\end{Definition}
\begin{Definition}{{Cycle}}
	A cycle $c$ in digraph $\Gamma(T)$ is the walk $(x_1,x_2)(x_2,x_3)\cdots(x_{k-1},x_k)$, such that $x_k = x_1$.
\end{Definition}
\begin{Definition}{{Maximum Cycle Mean}}\label{def:mcm}
	The maximum cycle mean, $\rho_\text{max} (T)$ is the maximum of the weight-to-length ratio of all cycles $c$ in $\Gamma(T)$, i.e.,
	\mr{
	\begin{equation}
	\label{eq:mcm}
	\footnotesize \rho_\text{max} (T) = \max\limits_{\forall c \text{ in }\Gamma(T)}\frac{|c|_T}{|c|} = \max\limits_{x_1,\cdots,x_{k-1}} \frac{T_{x_1 x_2} + \cdots T_{x_{k-1} x_k}}{k-1}
	\end{equation}
	}
\end{Definition}

Throughput of an SDFG is measured as the inverse of its \textit{maximum cycle mean} (Equation~\ref{eq:mcm}), i.e.,
\begin{equation}
    \label{eq:perf_def}
    \footnotesize \text{Performance (throughput)} = \frac{1}{\rho_\text{max} (T)}
\end{equation}

With this background, we now introduce \tech{}.

\begin{figure*}[h!]
	\centering
	\centerline{\includegraphics[width=1.99\columnwidth]{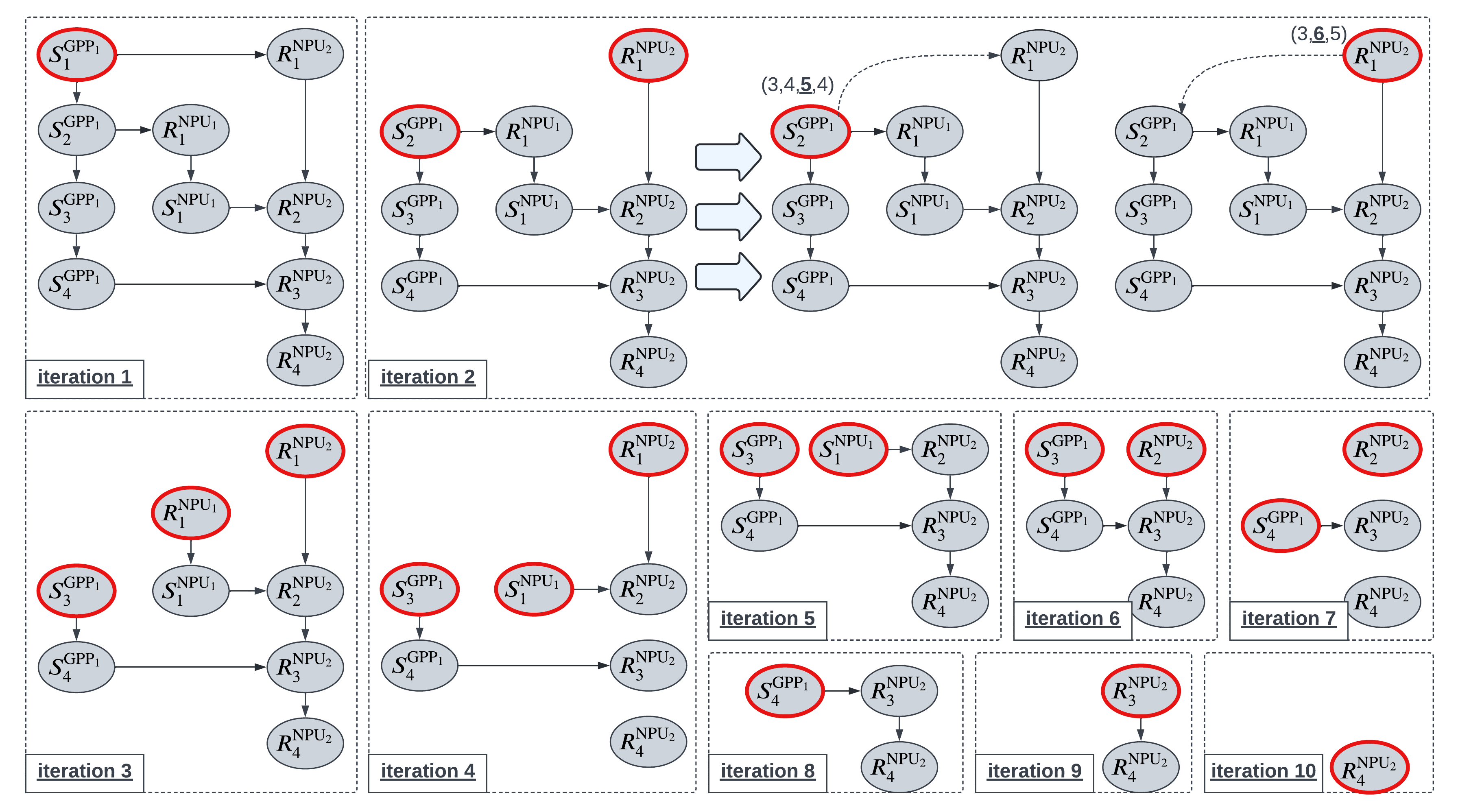}}
	\caption{Iterations of the transaction partial order algorithm to generate the transaction order of communication actors from the TPO graph of Figure~\ref{fig:ipc}d. Operations in each iteration is demonstrated using iteration 2 as an examples. First, ready actors are identified. Next, MCMs are computed for each ready actor by creating channels to other ready actors. The ready actor with the smallest MCM is selected as the candidate and deleted from the graph.}
	\label{fig:tpo}
\end{figure*}

\section{\tech{}: A \underline{P}erformance Oriented \underline{R}eal-T\underline{i}me \underline{S}cheduler for \underline{M}achine Learning Operations}\label{sec:prism}
\tech{} operates in three steps. 
First, it creates an \ipc{ (IPC)} graph from a machine learning application using the mapping of its operations on an \nsoc{'s} resources and a self-timed schedule.
Next, it estimates the throughput by embedding a transaction order for the inter-processor communication into the IPC graph and applying the maximum cycle mean formulation (Definition~\ref{def:mcm}). 
Finally, it explores the design space of scheduling operations to resources using a Hill Climbing heuristic to maximize the throughput.
We now discuss these steps in details.

\subsection{Creating IPC Graph}\label{sec:ipc}
\tech{} uses SDFGs to model an IPC graph. We illustrate this using the example of an inception block shown in Figure~\ref{fig:ipc}a. This is the building block of our evaluated InceptionV3 machine learning model~\cite{inception}.
There are 9 operations (called actors) -- input, conv 1-6, pool, and concat.
\tech{} uses the self-timed execution, where each computing resource executes the actors assigned to it in a fixed order that is specified at compile time.
Before firing an actor, it must wait for all of the actor's tokens to be available. We illustrate a self-timed schedule in Figure~\ref{fig:ipc}b for the specific actor allocation where input, conv 1, 3, \& 4 are mapped on GPP 1, conv 2 \& 3 on NPU 1, and pool, conv 6, and concat on NPU 2. 
To use self-timed strategy~\cite{SB00}, \tech{} enforces the actor execution order for each resource.
For instance, GPP 1 executes input, conv 3, conv 1, and conv 4 in this same order every time the inception block (i.e., its SDFG) is executed. 

Figure~\ref{fig:ipc}c shows the IPC graph that \tech{} constructs for the inception block.
Observe that the graph consists of computing actors, which represent operations of the inception block and additionally, a few extra actors which are shown with letters \texttt{S} and \texttt{R}. They
represent send and receive communication actors, respectively. 
\tech{} categorizes each channel of an SDFG as intra- or inter-processor channel, where a channel is called \textit{intra-processor channel} if its source and destination actors are mapped to the same resource, and \textit{inter-processor channel} otherwise.
In Figure~\ref{fig:ipc}b, there are 6 intra-processor and 5 inter-processor channels.
For each inter-processor channel, \tech{} adds two communications actors -- a send actor at its source and a receive actor at its destination. In Figure~\ref{fig:ipc}b, there are 5 send actors and 5 receive actors.

\subsection{Embedding A Transaction Order in an IPC Graph}\label{sec:perf}
For an \nsoc{}, inter-processor communications take place via the platform's shared resources.
So, a transaction order must be created for the communication actors. 
\tech{} uses the transaction partial order (TPO) graph, which it generates from an IPC graph by eliminating its computing actors~\cite{bambha2002intermediate}.
Figure~\ref{fig:ipc}d is the TPO graph of the IPC graph of Figure~\ref{fig:ipc}c.

\begin{figure*}[h!]
	\centering
	\centerline{\includegraphics[width=1.99\columnwidth]{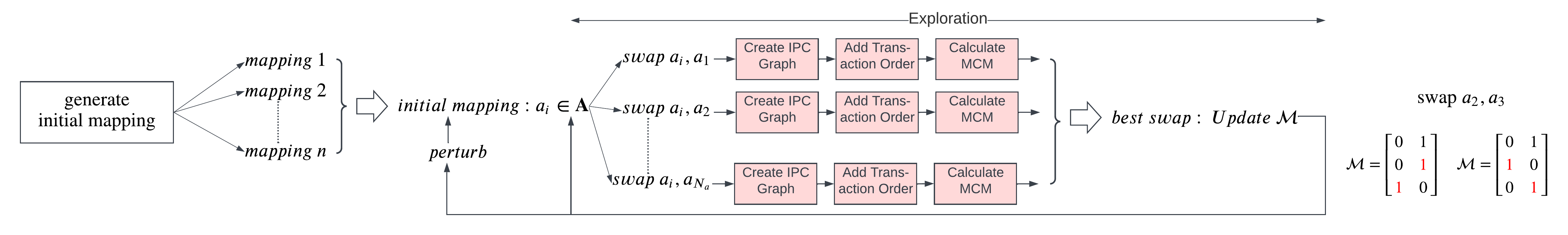}}
	\caption{Overview of the Hill Climbing heuristic to generate mapping of operations to computing units of an \nsoc{} platform. The swapping procedure is demonstrated to the right of this figure.}
	\label{fig:hill_climbing_demo}
\end{figure*}

Figure~\ref{fig:tpo} illustrates the iterations of a transaction partial order algorithm 
to generate the transaction order of communication actors .
Algorithm~\ref{alg:tpo_creation} shows its pseudo-code.
For each iteration, the algorithm performs the following steps.
First, it prepares a list of ready actors (i.e., those that do not have any incoming channel). 
In Figure~\ref{fig:tpo}, ready actors are \ineq{S_1^{\text{GPP} 1}} for iteration 1, \ineq{S_2^{\text{GPP} 1}} \& \ineq{R_1^{\text{NPU} 2}} for iteration 2, \ineq{S_3^{\text{GPP} 1}}, \ineq{R_1^{\text{NPU} 1}}, \& \ineq{R_1^{\text{NPU} 2}} for iteration 3, and so on.
Next, for each ready actor, it creates new outgoing channels to other ready actors. We illustrate these channels using dashed lines for iteration 2.
Next, it evaluates the MCM for these ready actors using Equation~\ref{eq:mcm}.
For iteration 2, we underline the MCMs out of other cycle means.
It calls a ready actor with the smallest MCM as \underline{candidate} and deletes it from the TPO graph along with all its outgoing channels.
In iteration 2, actor \ineq{S_2^{\text{GPP} 1}} is the candidate as it has the smallest MCM. 
The current iteration completes once the candidate is deleted from the TPO graph.
Subsequently, the algorithm repeats these steps for the next iteration, eventually terminating when all actors are deleted from the TPO graph.
Finally, the algorithm generates a transaction order by connecting all candidates in the sequence in which they are deleted.
Figure~\ref{fig:ipc}e shows the IPC graph with the transaction order of its communication actors shown using red dashed lines. 
\tech{} uses this IPC graph with embedded transaction order to compute the throughput using Equation~\ref{eq:perf_def}.


\subsection{Scheduling using a Hill Climbing Heuristic}\label{sec:scheduling}
Scheduling SDCNN operations on an \nsoc{} consists of determining the mapping, ordering, and timing~\cite{SB00}. Since \tech{} uses self-timed execution, it is not necessary to obtain the precise timing information. Here, we discuss how \tech{} explores the design space of mapping and ordering.

We introduce the following notations.

\begin{footnotesize}
\begin{align*}
G_\text{IPC}(A,C) &=~\text{An IPC graph with embedded transaction order}\\
A &=~\text{Set of computing and communication actors and } |A| = N_A\\
C &=~\text{Set of channels between actors}\\
G_{\nsoc{}}(R,E) &=~\text{An NSoC platform graph}\\
R &=~\text{Set of resources of the \nsoc{} and } |R| = N_R\\
E &=~\text{Set of edges/links between the resources}\\
\mathcal{M} &=~\{x_{i,j}\} =~\text{Mapping of }G_\text{IPC} \text{ on } G_{\nsoc{}} \\
x_{i,j} &= \begin{cases}
1 & \text{if actor } a_i\in {A} \text{ is mapped to resource } r_j\in R\\
0 & \text{otherwise}
\end{cases}
\end{align*}
\end{footnotesize}
\normalsize



The problem of finding a mapping of actors to resources draws parallel to mapping tasks on parallel computers~\cite{singh2013mapping}. 
This problem has been shown to be NP-complete for non-trivial optimization objectives~\cite{rosenberg1978data}.
Therefore, heuristic solutions such as Hill Climbing, Simulated Annealing, and Genetic Algorithms are effective in finding a solution~\cite{talbi1993hill}.
We use a Hill Climbing heuristic because of its simplicity~\cite{smitley1988comparative}.

\vspace{-10pt}
\begin{algorithm}[h]
	\scriptsize{
 		\KwIn{TPO graph ${G_{TPO}(\mathbf{V},\mathbf{E})}$}
 		\KwOut{Transaction order $\mathbf{O}$}
 		$\mathbf{O} = \emptyset$\tcc*[r]{Initialize the transaction order.}
 		\While(\tcc*[f]{Start of an iteration}){$\mathbf{V}\ne\emptyset$}{
 		    $\mathbf{R} = \{v\in\mathbf{V} | inC(v) = \emptyset\}$\tcc*[r]{Set of ready actors.}
 		    \For{$r\in\mathbf{R}$} {
 		        Create $Ch(r,u)~\forall u\in\mathbf{R} \text{ and } u\ne r$\tcc*[r]{Create a new channel from each ready actor to other ready actors.}
 		        $L_a.\texttt{append}(r)$ and $L_m.\texttt{append}\big(\texttt{ComputeMCM}(G_{TPO})\big)$\tcc*[r]{Store the ready actor and the corresponding MCM in local variables $L_a$ and $L_m$, respectively.}
 		        $x = \texttt{argmin}(L_m)$\tcc*[r]{Find the index to the minimum MCM.}
 		        $\texttt{candidate} = L_a[x]$\tcc*[r]{Select an actor with the minimum MCM as candidate.}
 		        $\mathbf{O}.\texttt{append}(\texttt{candidate})$\tcc*[r]{Insert the \texttt{candidate} in the transaction order $\mathbf{O}$.}
 		        $\mathbf{V}\setminus \{\texttt{candidate}\}$ and $\mathbf{E}\setminus \{OutC(\texttt{candidate})\}$ \tcc*[r]{Delete the \texttt{candidate} and all its outgoing channels.}
 		    }
 		}
 		\textbf{return} $\mathbf{O}$
 	}
	\caption{\footnotesize{Transaction partial order algorithm.}}
	\label{alg:tpo_creation}
\end{algorithm}
\vspace{-10pt}

Figure~\ref{fig:hill_climbing_demo} illustrates the steps for the proposed Hill Climbing heuristic.
It starts with the \emph{generate initial mapping} block, 
which generates a set of initial mappings.
Next, it selects one of these starting mappings and proceeds to the exploration stage.
Here, it performs a trial swap for each actor \ineq{a_i\in \mathbf{A}} with another actor that is mapped to a different resource.
A trial swap operation involves changing the mapping matrix temporarily, as we illustrate to the right of the figure.
For each of these trial swaps, our heuristic evaluates the optimization objective -- the MCM. For this, we follow the procedure outlined in Sections~\ref{sec:ipc} \& \ref{sec:perf}, which involves (1) creating an IPC graph from the mapping obtained after performing a swap, (2) embedding it with a transaction order for communication actors using Alg.~\ref{alg:tpo_creation}, and (3) computing the MCM using Eq.~\ref{eq:mcm}. 

Next, the heuristic selects a swap with the minimum MCM because reducing the MCM increases its throughput.
In case of a tie, it selects a trial swap randomly.
Once a swap is finalized, it makes the swap permanent by updating the mapping matrix.
It then iterates through these steps for the next actor.
A \emph{pass} in this heuristic involves
completing trial swaps for every pair of actors mapped to different resources.
If the objective function improves during a pass, we initiate another pass of our heuristic.
Otherwise, we call the heuristic to be stuck at a local minimum.
To come out of this local minimum, we perturb the current mapping and restart the exploration, where perturbing involves performing a fixed number of random swaps. 

During the Hill Climbing heuristic, we record all mappings, i.e., the initial mapping and the mappings obtained after performing each swap operation. For all these mappings, we estimate the energy consumption of their corresponding schedule. A set of Pareto-optimal mappings are only retained for use-case mapping, which is described next.


\section{Mapping Multi-Application Use-Cases on \nsoc{}}\label{sec:usecases}
\begin{figure*}[h!]
	\centering
	\centerline{\includegraphics[width=1.99\columnwidth]{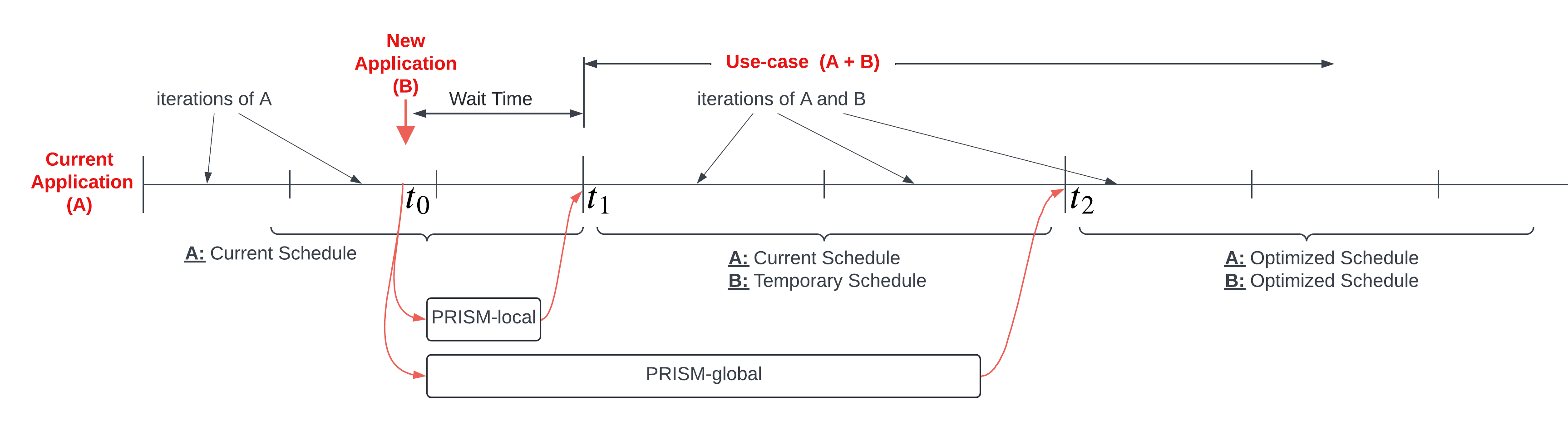}}
	\caption{Scheduling multi-application use-cases on an \nsoc{} via \tech{'s} local and global explorations.}
	\label{fig:usecase}
\end{figure*}


Figure~\ref{fig:usecase} illustrates how \tech{} creates schedules for more than one application at the same time. Imagine a user initiating application B at time \ineq{t_0}, when the platform is already executing application A. 
Therefore, operations of B must be scheduled on the same resources that are currently executing A. This is a huge undertaking and it involves guaranteeing performance for both these applications to deliver a quality-of-service.
One solution could be to analyze every combination of applications at design time and store the corresponding schedule -- with \ineq{n} applications, there are \ineq{n^2} use-cases to analyze. However, storing all use-cases from design-time can increase the storage overhead of an embedded platform and provides limited flexibility for run-time adaptation.

To address this, \tech{} initiates a local and global exploration. 
The objective of \tech{'s} local exploration is to create a quick schedule for B keeping the current schedule of A unchanged. This is to admit B in the shortest possible time while providing a certain performance guarantee and ensuring that the performance of A does not degrade excessively. In the figure, \tech{} admits B to the platform at time \ineq{t_1}. 

The objective of \tech{'s} global exploration is to find optimized schedules for both A and B, which provide the best system performance while sharing the resources of an \nsoc{}.
In the figure, \tech{} implements the optimized schedule for A and B from time \ineq{t_2} onwards.

\subsection{Local Exploration of \tech{}}
Algorithm~\ref{alg:local_prism} shows the pseudo-code of \tech{'s} local exploration. Here, \ineq{SchA} is the schedule of application A, i.e., the schedule currently running on the \nsoc{}. First, \tech{} retrieves all Pareto-optimal schedules of the newly enabled application B and store it in the set \ineq{S_B} (line 2). Next, for each schedule \ineq{Sch_B\in S_B}, it computes the resource contention related overhead using a probabilistic framework (line 4). Finally, it selects a schedule for B that minimizes this overhead. The key idea is to reduce the expected wait time before actors from the two applications can be scheduled on contending resources. \tech{} uses this new schedule for B after it finishes exploring all alternatives. In the mean time, the platform continues to execute the current schedule of A. In this way, \tech{} ensures a non-preemptive system where the actors that are already executing on the platform are not preempted to provide resources for the new application.

\vspace{-10pt}
\begin{algorithm}[h]
	\scriptsize{
 		\KwIn{Schedule database $sDB$, schedule $SchA$ of A, and application B}
 		\KwOut{Schedule $SchB$ of B.}
 		$\rho_A = MCM(SchA)$\tcc*[r]{MCM of A.}
 		$S_B = sDB(B)$\tcc*[r]{Pareto-optimal schedules of B.}
 		\For(\tcc*[f]{For each schedule in $S_B$}){$SchB\in S_B$} {
 		    $SchA,SchB = \texttt{Contention}(SchA,SchB)$\tcc*[r]{Compute resource sharing related contention.}
 		    $\hat{\rho}_A = MCM(SchA)$ and $\rho_B = MCM(SchB)$\tcc*[r]{Compute MCM of A and B considering resource contention.}
 		    $\Delta\rho_A = 100*\frac{\hat{\rho}_A-{\rho_A}}{{\rho_A}}$\tcc*[r]{Compute the performance degradation of A due to resource contention.}
 		    \If{ $\Delta\rho_A < \Delta\rho_{max}$ and $\rho_B < \rho_{constraint}$}{
 		        \textbf{return} $SchB$
 		    }
 		}
 	}
	\caption{\footnotesize{Local exploration of \tech{}.}}
	\label{alg:local_prism}
\end{algorithm}
\vspace{-10pt}

To formulate the resource contention related overhead, we consider that each actor must be in one of the three states at any given time -- (1) \underline{not-ready state}, when it waits for all of its tokens to be available, (2) \underline{wait state}, when it waits for its resource to be available, and (3) \underline{execution state}, when it is currently executing a machine learning operation. Now assume that actor \texttt{b} \ineq{\in} B is mapped on the same resource as actor \texttt{a} \ineq{\in} A. We are to model the expected execution time of \texttt{a} and \texttt{b} when contending for the resource.
When \texttt{b} is enabled, it can find \texttt{a} in one of the three states.
Let the random variable \ineq{S(t)} denote the state of actor \texttt{a} and \ineq{Y} denote the wait time of \texttt{b}.

We define the following probabilities.

\vspace{-10pt}
\begin{footnotesize}
\begin{eqnarray}
\label{eq:probabilities}
P\big(S(t) = S_w\big) &=& \text{probability of } \texttt{a} \text{ in wait state} \\
&=& t_{wait}\cdot \rho_{max} \nonumber \\
P\big(S(t) = S_e\big) &=& \text{probability of } \texttt{a} \text{ in execution state}\nonumber \\
&=& t_{a}\cdot \rho_{max} \nonumber \\
P\big(S(t) = S_n\big) &=& \text{probability of } \texttt{a} \text{ in not-ready state}\nonumber \\
&=& 1 - P\big(S(t) = S_w\big) - P\big(S(t) = S_e\big) \nonumber
\end{eqnarray}
\end{footnotesize}\normalsize
where \ineq{t_{wait}} is the average wait time and \ineq{\rho_{max}} is the MCM of the IPC of A. The above steady-state probabilities are derived assuming stationarity of an actor in each state. 

The expected wait time is obtained as
\begin{equation}
    \label{eq:ey}
    \footnotesize E(Y) = \int_{-\infty}^\infty y~P(y)~dy,
\end{equation}
where \ineq{P(y)} is the probability density function of \ineq{Y}. We make the following consideration in solving Equation~\ref{eq:ey}. First, if \texttt{b} arrives when \texttt{a} is in not-ready state, then \texttt{b} can be scheduled immediately. Second, if \texttt{b} arrives when \texttt{a} is in wait state (due to resource contention from other actors), then \texttt{b} must wait for the entire duration of the \texttt{a}'s execution time. In this case, \texttt{b} is pushed to the back of the ready queue for the resource. Here, we uses first-come-first-serve (FCFS) scheduling for each resource. Finally, if \texttt{b} arrives when \texttt{a} has started executing, then \texttt{b} must wait for the remaining time until \texttt{a} finishes execution. Therefore, Equation~\ref{eq:ey} can be rewritten as

\vspace{-10pt}
\begin{footnotesize}
\begin{eqnarray}
    \label{eq:ey2}
     E(Y) =&& y_{|_{S(t) = S_n}}\cdot P\big(S(t) = S_n\big)\\ 
     && +~y_{|_{S(t) = S_w}}\cdot P\big(S(t) = S_w\big) \nonumber \\ 
     && +~y_{|_{S(t) = S_e}}\cdot P\big(S(t) = S_e\big)\nonumber
\end{eqnarray}
\end{footnotesize}\normalsize

Assuming the remaining execution time of \texttt{a} is uniformly distributed within the time duration of the execution time of \texttt{a}, the above equation reduces to
\begin{equation}
    \label{eq:ey3}
    \footnotesize E(Y) = 0\cdot P\big(S(t) = S_n\big) + t_{a}\cdot t_{wait}\cdot \rho_{max} + \frac{t_a}{2}\cdot t_{a}\cdot \rho_{max}
\end{equation}

Since \ineq{E(Y)} is the average wait time \ineq{t_{wait}}, Equation~\ref{eq:ey3} can be solved to obtain
\begin{equation}
    \label{eq:wait_time}
    \footnotesize t_{wait} =  \frac{t_a^2\cdot \rho_{max}/2}{1-t_a\cdot\rho_{max}}
\end{equation}

Finally, the modified execution time of \texttt{b} is 
\begin{equation}
    \label{eq:modified_ex_time}
    \footnotesize t_b = t_b + t_{wait}
\end{equation}

This formulation needs to be extended for every actor of the two applications A and B. Once completed, Algorithm~\ref{alg:local_prism} computes the modified MCM for the two applications (line 5).
It estimates the performance degradation of A (line 6). If they both are acceptable, i.e., they are within the user defined limits, Algorithm~\ref{alg:local_prism} terminates, returning the new schedule of B (lines 7-8). Here, \ineq{\Delta\rho_{max}} is the maximum acceptable throughput degradation of A and \ineq{\rho_{constraint}} is the throughput constraint of B. These are user-defined parameters.

\subsection{Global Exploration of \tech{}}
For global exploration, \tech{} first merges the IPC graphs of the two application. 
Let \ineq{G_A(A_1,C_1)} be the IPC graph of application A with the set \ineq{A_1} of actors and set \ineq{C_1} of channels. 
Let \ineq{G_B(A_2,C_2)} be the IPC graph of application B with the set \ineq{A_2} of actors and set \ineq{C_2} of channels. 
The merged graph is 
\begin{equation}
    \label{eq:merged_a_b}
    \footnotesize G_{AB}(A,C)~~|~~ A = A_1 \cup A_2 \text{ and } C = C_1 \cup C_2
\end{equation}

Next, the merged graph is analyzed using the formulation that we presented in Section~\ref{sec:performance}. New schedules for A and B are implemented after completing their ongoing iterations.



\section{Evaluation Methodology}\label{sec:evaluation}
\subsection{Simulation Framework}
We implement \tech{} inside NeuroXplorer~\cite{icons2021}, an architectural simulator of neuromorphic system-on-chip platforms.
We perform all simulations on a Lambda workstation, which has AMD Threadripper 3960X with 24 cores, 128 MB cache, 128 GB RAM, and 2 RTX3090 GPUs. Table~\ref{tab:hw_parameters} shows our simulation parameters. The code is available online at {https://github.com/drexel-DISCO/PRISM}.

\begin{table}[h!]
    \renewcommand{\arraystretch}{1.0}
    \caption{Major simulation parameters.}
	\label{tab:hw_parameters}
	\vspace{-5pt}
	\centering
	{\fontsize{6}{10}\selectfont
		\begin{tabular}{lp{5cm}}
			\hline
			Number of GPPs & 2\\
			\hline
			Number of NPUs & 128\\
			\hline
			\nsoc{} Platform & $\mu$Brain~\cite{chen1999segmented}, SPECK~\cite{speck},\& GrAI~\cite{grai}\\
			\hline
			Design Parameters & Energy~\cite{cf2021}, Throughput~\cite{lctes2020}, Reliability~\cite{mwscas2020,tpds2021}, and Technology~\cite{dt2022}\\
			\hline
			NPU supported operations -- $\mu$Brain & Activation, Add, Conv2D, Concatenate, Dense, InputLayer, Normalization \\
			NPU supported operations -- SPECK & Activation, Add, AveragePooling2D, Concatenate, Conv2D, Dense, Dropout, Flatten, GlobalAveragePooling2D, GlobalMaxPooling2D, InputLayer, MaxPooling2D, Normalization, ReLU\\
			NPU supported operations -- GrAI & BatchNormalization, ZeroPadding2D, DepthwiseConv2D, Reshape, Rescaling, Multiply, and all supported operations of SPECK\\
			\hline
	\end{tabular}}
\end{table}

\begin{table}[h!]
	\renewcommand{\arraystretch}{1.0}
	\setlength{\tabcolsep}{1.0pt}
	\caption{Evaluated models and ImageNet Top-1 accuracy.}
	\label{tab:apps}
	\centering
	\begin{threeparttable}
	{\fontsize{6}{10}\selectfont
		\begin{tabular}{cc|cc|cc|cc}
			\hline
			\textbf{Models} & \textbf{Accuracy} & \textbf{Models} & \textbf{Accuracy} & \textbf{Models} & \textbf{Accuracy} & \textbf{Models} & \textbf{Accuracy} \\
			\hline
			LeNet	&	49.0\% &	ResNet50	&	77.1\%	&	DenseNet121	&	74.4\%	&	MobeleNet	&	74.2\%	\\
            AlexNet	&	60.5\%	&	ResNet101	&	78.0\%	&	DenseNet169	&	76.1\%	&	ShuffleNet	&	70.8\%	\\
            ZFNet	&	64.0\%	&	InceptionV3	&	78.1\%	&	SqueezeNet	&	57.5\%	&	NASNetMobile	&	77.9\%	\\
            VGG16	&	73.4\%	&	ResNext	&	83.5\%	&	Xception	&	78.6\%	&	NoisyStudent	&	82.0\%	\\
			\hline
	\end{tabular}}
	\end{threeparttable}
\end{table}


\subsection{Evaluated Applications}
We evaluate \tech{} using 20 models -- 16 models trained on the ImageNet dataset and four additional state-of-the-art models. The ImageNet models are summarized in Table~\ref{tab:apps}. Although improving accuracy is not the focus here, we provide accuracy numbers for completeness.
The four additional models are (1) PilotNet~\cite{bojarski2020nvidia} for self-driving cars, (2) the transformer-based BERT~\cite{bert} for natural language processing, (3) U-Net~\cite{unet} for medical image segmentation, and (4) ConvLSTM~\cite{convlstm} for time-series data processing.


We use the following use-cases for evaluating the local and global explorations of \tech{} for scheduling use-cases.
\begin{enumerate}
    \item \textbf{Usecase-1:} A combination of BERT and MobileNet
    \item \textbf{Usecase-2:} A combination of ResNet50 and U-Net
    \item \textbf{Usecase-3:} A combination of SqueezeNet and AlexNet
    \item \textbf{Usecase-4:} A combination of VGG16 and ShuffleNet
    \item \textbf{Usecase-5:} A combination of Xception and PilotNet
\end{enumerate}



\subsection{Models Supported on Evaluated \nsoc{s}}
\begin{figure}[h!]
	\centering
	\vspace{-10pt}
	\centerline{\includegraphics[width=0.99\columnwidth]{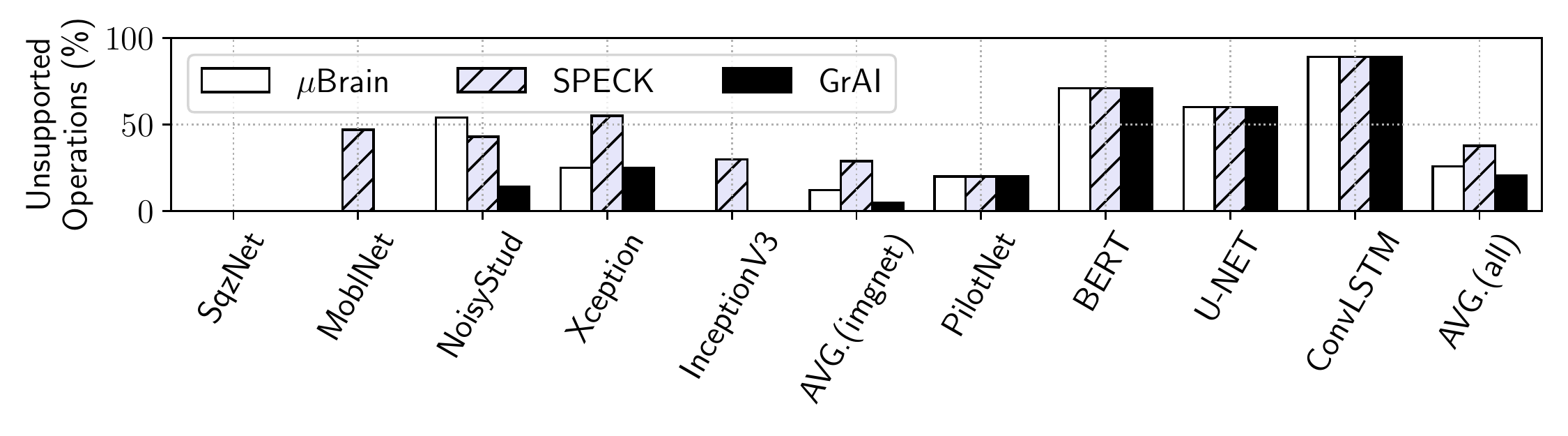}}
	\vspace{-10pt}
	\caption{Unsupported operations on three state-of-the-art \nsoc{} platforms.}
	\vspace{-5pt}
	\label{fig:unsupported_operations}
\end{figure}

As we discuss in Section~\ref{sec:introduction}, an NPU may not support all operations of a machine learning model. To give further insight, Figure~\ref{fig:unsupported_operations} shows the fraction of unsupported operations for three state-of-the-art \nsoc{s} -- $\mu$Brain~\cite{date2022}, SPECK~\cite{speck}, and GrAI~\cite{grai}. 
We report results for 5 ImageNet models and average across 16 such models of Table~\ref{tab:apps}. We also report results for the four additional models and the average across all 20 evaluated models.
We make the following observations.

Some models such as SqueezeNet (abbreviated as SqzNet) is supported on all three platforms (unsupported operations is close to 0\%). This is because this model consists of standard CNN operations such as convolution, pooling, and dense, which are all supported on these three platforms. On the other hand, most operations of BERT, U-Net and ConvLSTM are not supported on any of these platforms due to the use of specialized operations. For ImageNet models, $\mu$Brain, SPECK, and GrAI do not support on average 12\%, 30\%, and 5\% of the total operations, respectively. Considering all models, these fractions are 26\%, 38\%, and 21\%, respectively. \tech{} uses the GPPs to schedule all unsupported operations.

\subsection{Evaluated Schedulers}
We evaluate the following schedulers.
\begin{enumerate}
    \item \textbf{Baseline~\cite{tecs2021}:} This is an NPU-only policy. 
    It uses a GPP to schedule only the operations not supported on the NPU.
    It exploits only batch parallelism.
    \item \textbf{SentryOS~\cite{date2022}:} This is  the baseline that exploits both pipeline and batch parallelism. 
    \item \textbf{\tech{}}: The is the proposed scheduler which exploits platform heterogeneity to schedule operations. Here, an operation can be scheduled on a GPP even if it is supported on an NPU. Using a Hill Climbing heuristic, \tech{} creates opportunities for all three forms of parallelism -- batch, pipeline, and operation.
\end{enumerate}

\section{Results}\label{sec:results}
\subsection{Throughput Performance}\label{sec:speedup_results}
\begin{figure}[h!]
	\centering
	\vspace{-10pt}
	\centerline{\includegraphics[width=0.99\columnwidth]{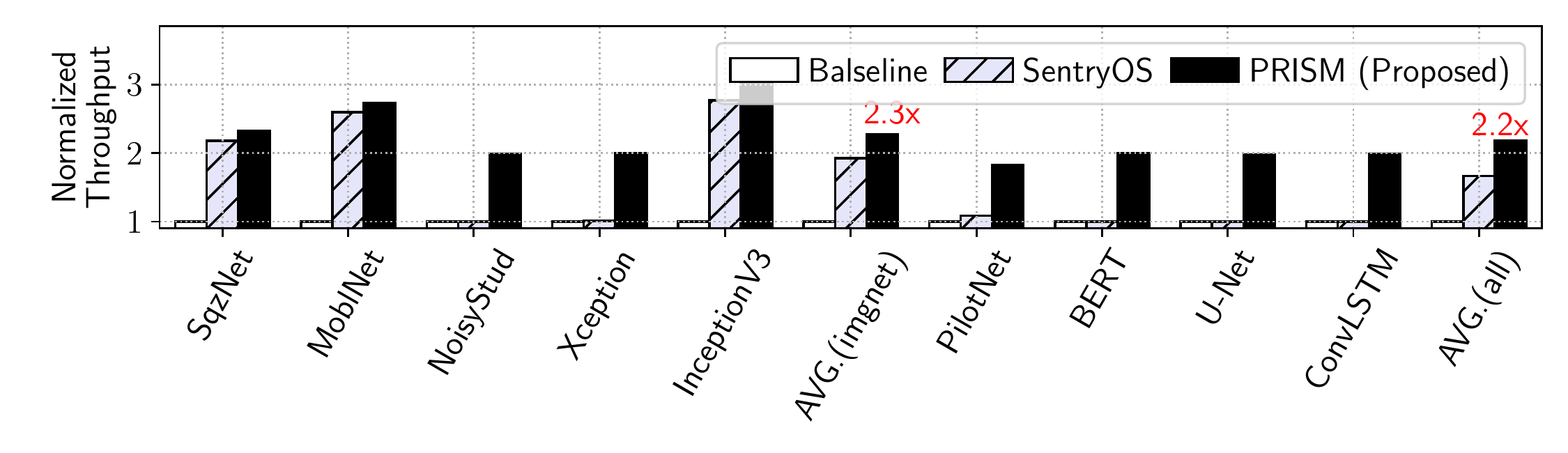}}
	\vspace{-10pt}
	\caption{Throughput normalized to Baseline.}
	\vspace{-5pt}
	\label{fig:speedup}
\end{figure}

Figure~\ref{fig:speedup} reports throughput of the evaluated schedulers 
We make the following observations.

Although both \base{} and \pb{} exploit 
batch parallelism, \pb{} additionally exploits pipeline parallelism within each batch. So, the throughput of \pb{} is higher (on average, 1.9x higher for ImageNet models and 1.6x higher for all 20 models). 
Between \pb{} and the proposed \tech{}, \pb{} uses a GPP only for those operations that are not supported on an NPU.
Therefore, the degree of pipeline and operation parallelism that can be exploited is limited.
On the other hand, \tech{} has a higher degree of freedom in mapping operations to GPPs and NPUs. So, \tech{} can better exploit these forms of parallelism using the Hill Climbing heuristic. 
For ImageNet models, \tech{} has 2.3x and 1.2x higher throughput than \base{} and \pb{}, respectively. Considering all 20 models, the throughput of \tech{} is 2.2x and 1.3x higher, respectively.

\subsection{Performance Per Watt}\label{sec:sppedup_per_joule}
\begin{figure}[h!]
	\centering
	\vspace{-10pt}
	\centerline{\includegraphics[width=0.99\columnwidth]{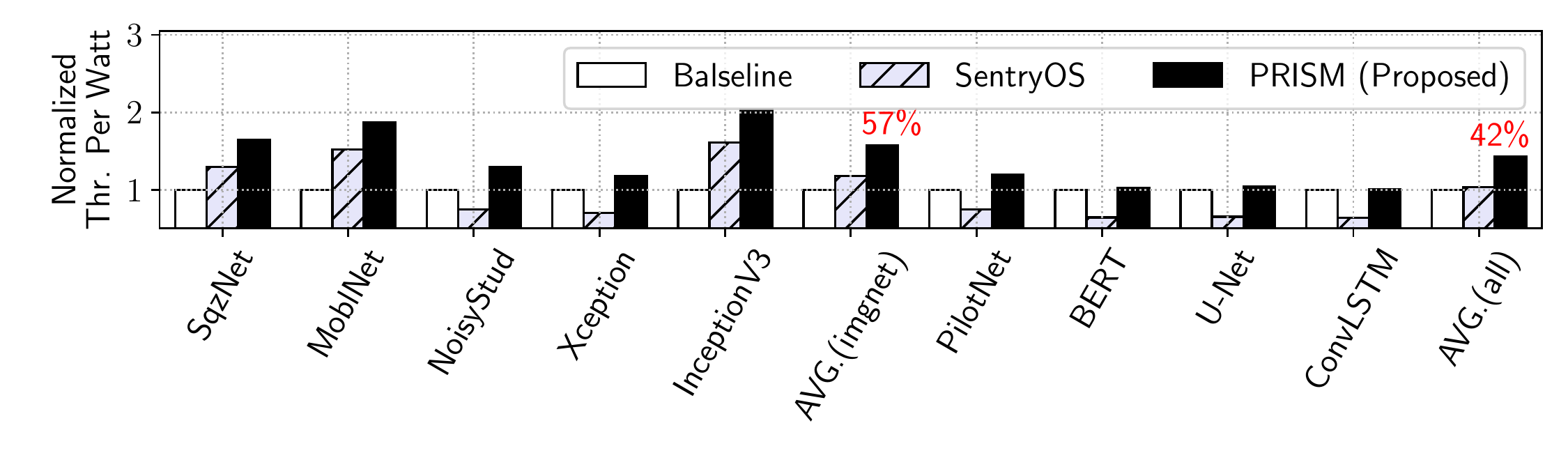}}
	\vspace{-10pt}
	\caption{Throughput per Watt normalized to Baseline (higher is better).}
	\vspace{-5pt}
	\label{fig:perf_joule}
\end{figure}

Figure~\ref{fig:perf_joule} reports the throughput per watt of the evaluated schedulers. We make the following observations.

Between  \base{} and \pb{}, \pb{} has 18\% higher throughput per watt on average than \base{}. Although \pb{} has 1.9x higher speedup than \base{} (see Section~\ref{sec:speedup_results}), it also has a higher energy consumption due to an increased utilization of resources in exploiting pipeline parallelism. For the 4 non-ImageNet models where the throughput improvement is relatively small, the throughput per watt is lower than \base{}. Considering all 20 models, \pb{} has only 3\% higher throughput per watt compared to \base{}.

Between \pb{} and \tech{}, \tech{} has a higher throughput (average 1.3x higher) and lower energy (average 5\% lower). So, the throughput per watt is 33\% higher than \pb{} for ImageNet models and 38\% higher for all 20 models. Compared to \base{}, \tech{} has 57\% and 42\% higher throughput per watt, respectively.

\subsection{Energy Breakdown}\label{sec:energy_breakdown}
\begin{figure}[h!]
	\centering
	\vspace{-10pt}
	\centerline{\includegraphics[width=0.99\columnwidth]{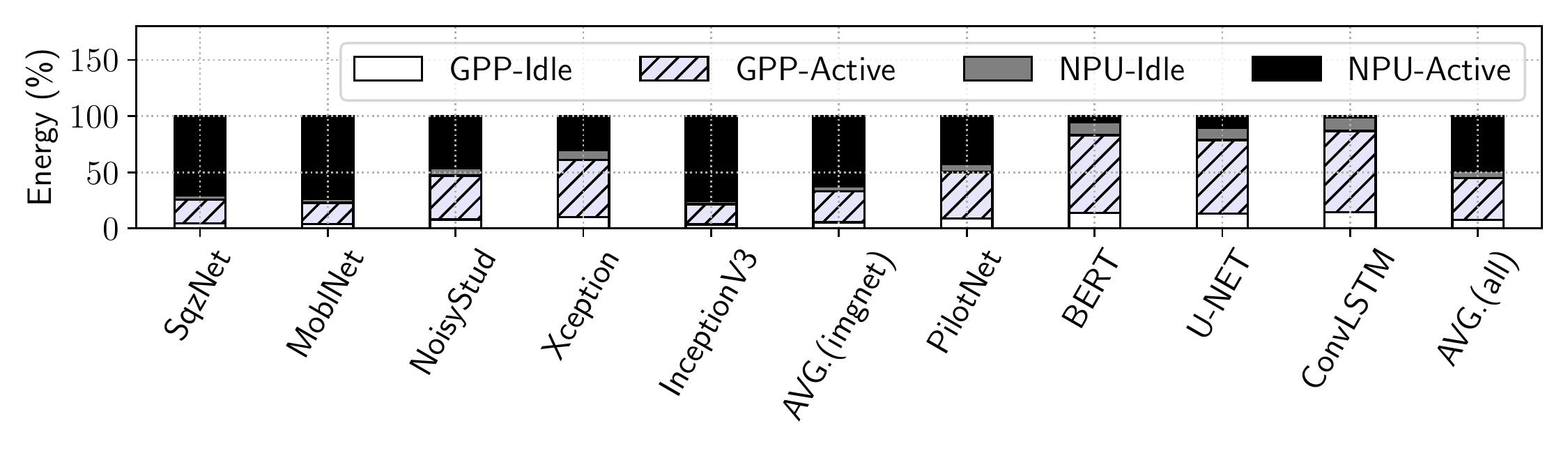}}
	\vspace{-10pt}
	\caption{Total energy distributed into GPP and NPU energy (idle and active).}
	\vspace{-5pt}
	\label{fig:energy_breakdown}
\end{figure}

Figure~\ref{fig:energy_breakdown} reports the total energy of each model, distributed into GPP and NPU energy. The figure reports both active and idle energy, where active energy is the energy consumed when a resource is performing an operation, and idle energy is the energy consumed while it is not performing any operation. We make the following observations.

When idle, resources still draw a portion of the peak power.
Therefore, the idle GPP and NPU energy constitute on average 8\% and 6\% of the total energy, respectively.
Even though all operations of SqueezeNet are supported on an NPU (see Figure~\ref{fig:unsupported_operations}), GPP active energy constitutes about 21\% of the total energy. This is because \tech{} schedules its operations using both GPPs and NPUs to improve the throughput. 
For BERT, U-Net, and ConvLSTM, \tech{} uses the GPP to schedule most operations.
So, 
the GPP active energy is the dominant component of the total energy (on average 69\%).

\subsection{Pareto Exploration during Hill Climbing Optimization}
\begin{figure}[h!]%
    \centering
    \vspace{-15pt}
    \subfloat[SqueezeNet.\label{fig:squeezenet}]{{\includegraphics[width=2.9cm]{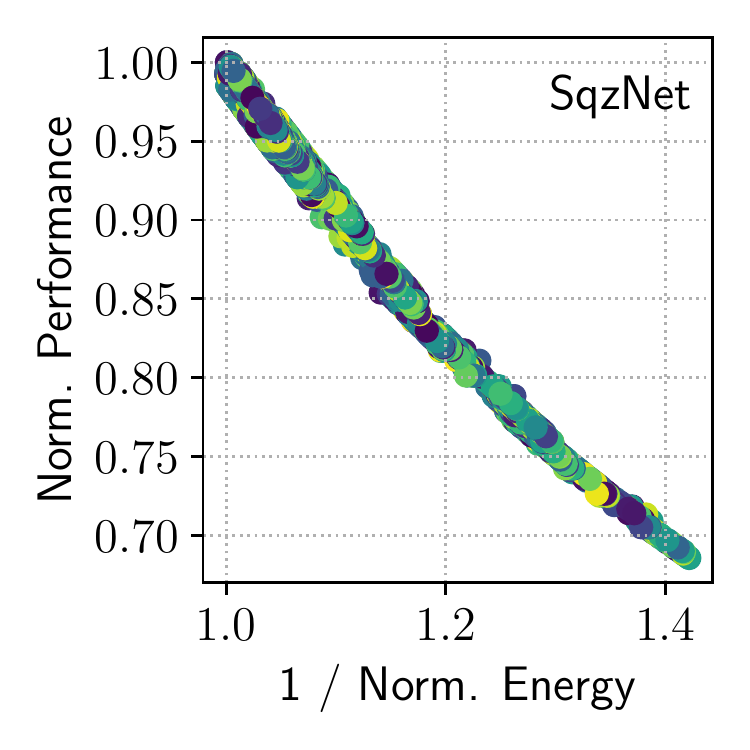} }}%
    \hfill
    \subfloat[MobileNetV3.\label{fig:mobilenet}]{{\includegraphics[width=2.9cm]{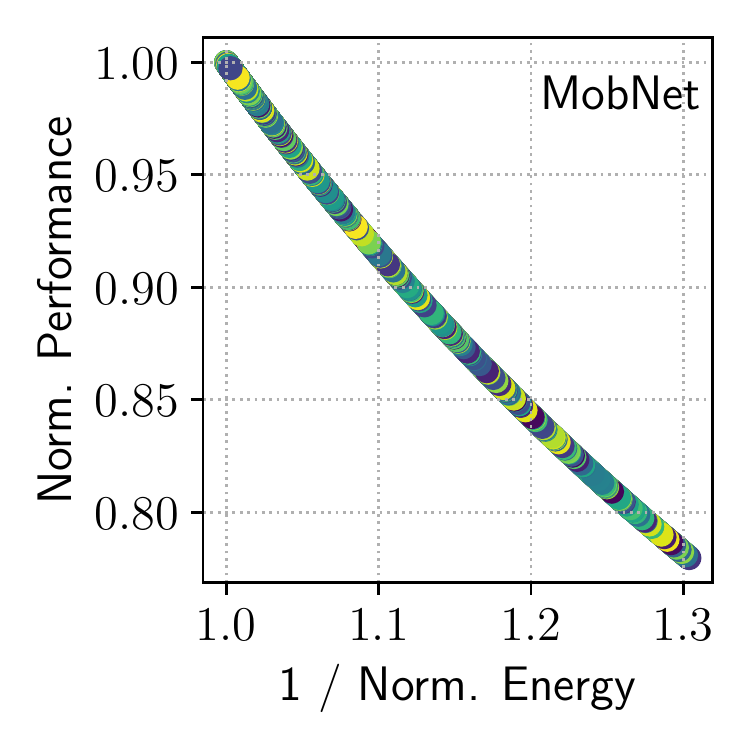}}}%
    \hfill
    \subfloat[Xception.\label{fig:xception}]{{\includegraphics[width=2.9cm]{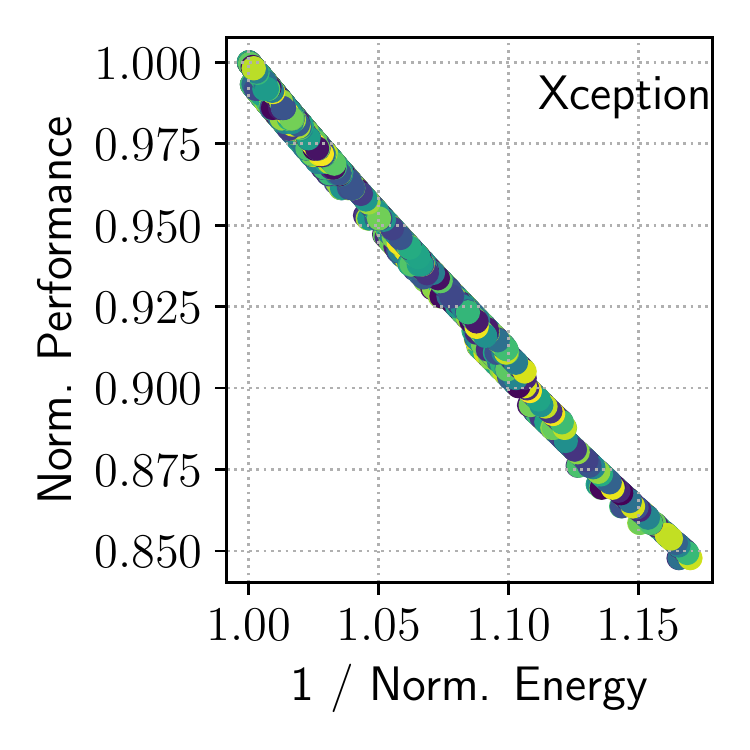}}}%
    \hfill
    \subfloat[PilotNet.\label{fig:pilotnet}]{{\includegraphics[width=2.9cm]{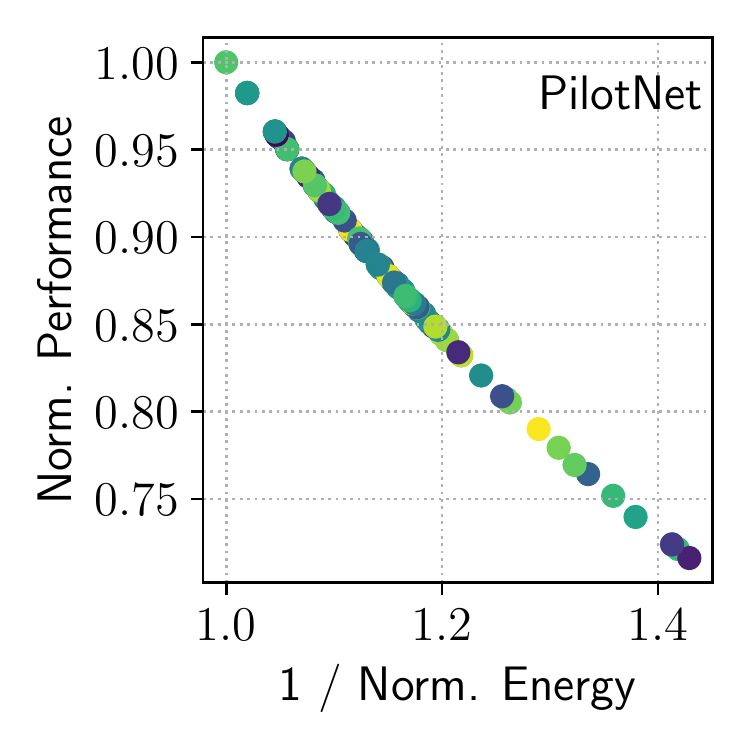} }}%
    \hfill
    \subfloat[BERT.\label{fig:bert}]{{\includegraphics[width=2.9cm]{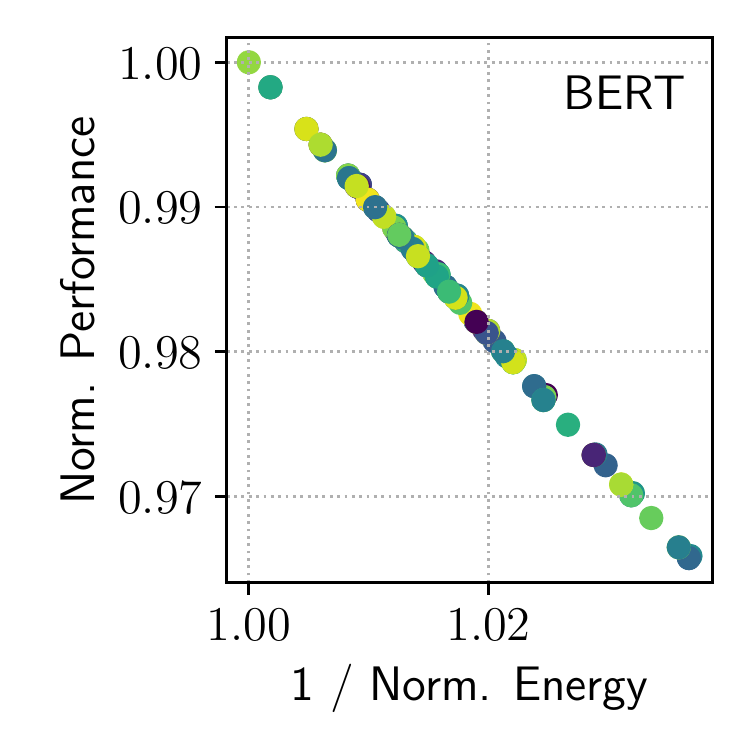}}}%
    \hfill
    \subfloat[U-Net.\label{fig:unet}]{{\includegraphics[width=2.9cm]{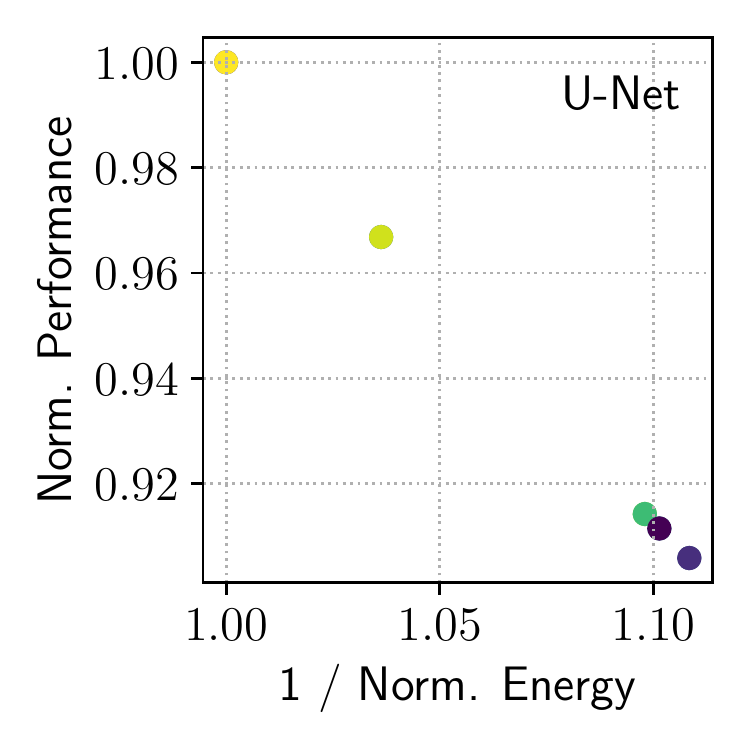}}}%
    \caption{Pareto points retained during the Hill Climbing exploration.}%
    \label{fig:hco_exploration}%
\end{figure}

Figure~\ref{fig:hco_exploration} shows the energy-throughput Pareto points retained during the proposed Hill Climbing based mapping exploration for six models. \tech{} selects the highest throughput point for every model for a given energy constraint.

\vspace{-10pt}

\subsection{Throughput Scalability}
\begin{figure}[h!]%
    \centering
    \subfloat[Throughput normalized to Baseline across evaluated platforms.\label{fig:platform_scalability}]{{\includegraphics[width=8.5cm]{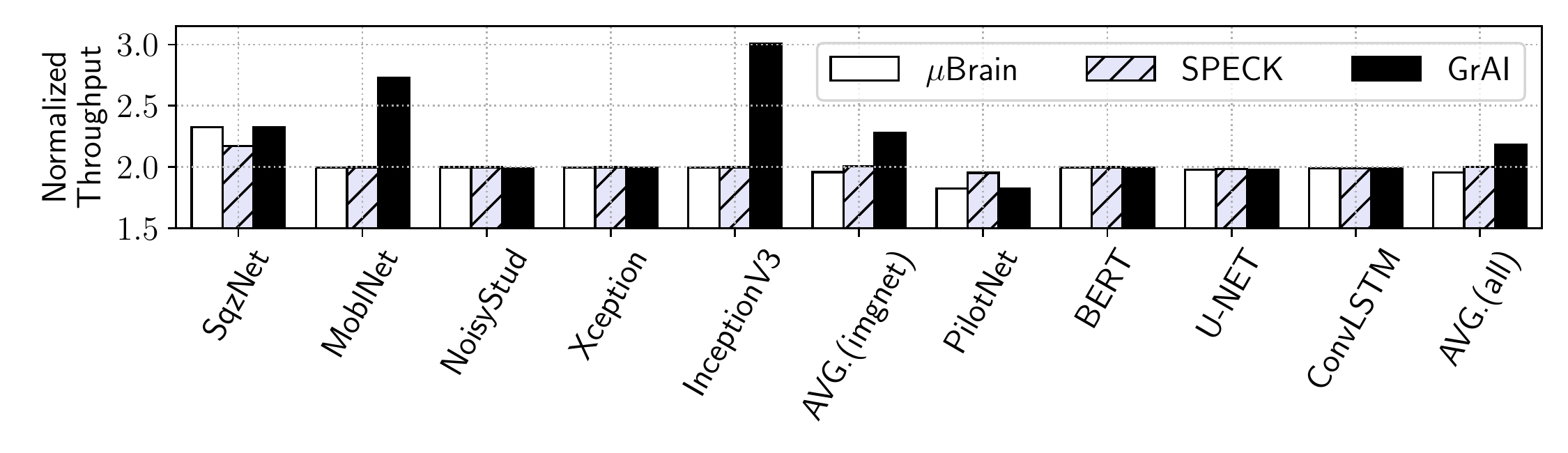} }}%
    \vspace{-10pt}
    \hfill
    \subfloat[Throughput normalized to Baseline with increasing number of NPUs.\label{fig:npu_scalability}]{{\includegraphics[width=8.5cm]{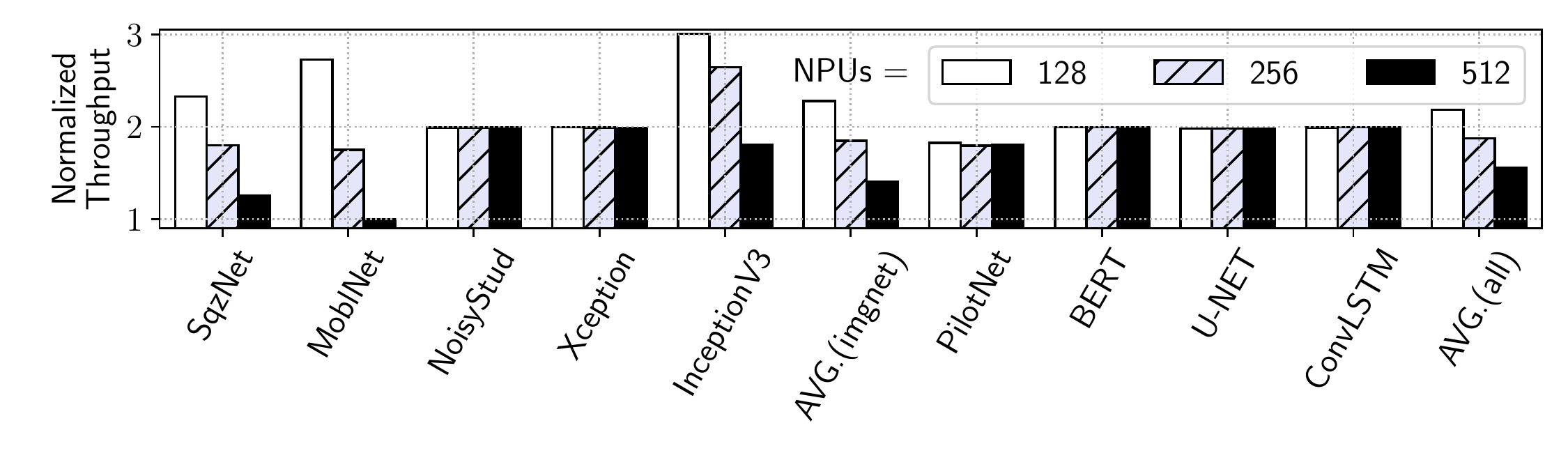}}}%
    \vspace{-10pt}
    \hfill
    \subfloat[Throughput normalized to Baseline with increasing number of GPPs.\label{fig:gpp_scalability}]{{\includegraphics[width=8.5cm]{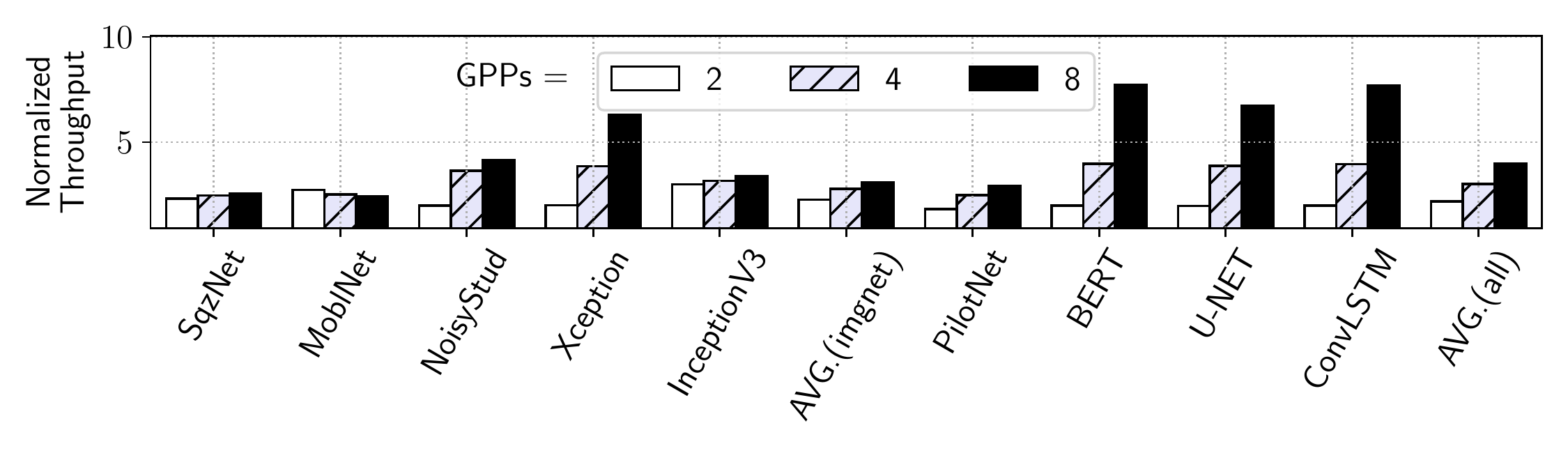}}}%
    \caption{Speedup of \tech{} for (a) different evaluated platforms, (b)~increasing number of NPUs, and (c) increasing number of GPPs.}%
    \label{fig:scalability}%
\end{figure}

Figure~\ref{fig:platform_scalability} reports the throughput of \tech{} normalized to \base{} for the three evaluated \nsoc{s}. For BERT, U-Net, and ConvLSTM, the throughput is comparable. This is because most operations of these three models are not supported on an NPU. Therefore, \tech{} uses GPPs for these operations, which results in similar performance across the three \nsoc{} platforms. For other applications, throughput is higher for the platform where the heterogeneity can be exploited better. 

Figure~\ref{fig:npu_scalability} reports the throughput of \tech{} normalized to \base{} as we increase the number of NPUs from 128 (base config) to 512. We observe that the relative throughput reduces due to this change. This is because with more NPUs, batch parallelism is the dominant factor that contributes to performance. Since both \tech{} and \base{} exploits this parallelism, the relative performance between these two schedulers reduces. \tech{} is still better because of the pipeline and operation parallelism that it additionally exploits from the hardware.

\vspace{-10pt}
\vspace{-10pt}

Figure~\ref{fig:gpp_scalability} reports the throughput of \tech{} normalized to \base{} as we increase the number of GPPs from 2 (base config) to 8. We observe that the relative throughput improves due to this change. This is because with more GPPs, \tech{} can better exploit operation parallelism in an \nsoc{}, which improves the throughput.


\subsection{Use-case Performance}
\begin{figure}[h!]
	\centering
	\vspace{-10pt}
	\centerline{\includegraphics[width=0.99\columnwidth]{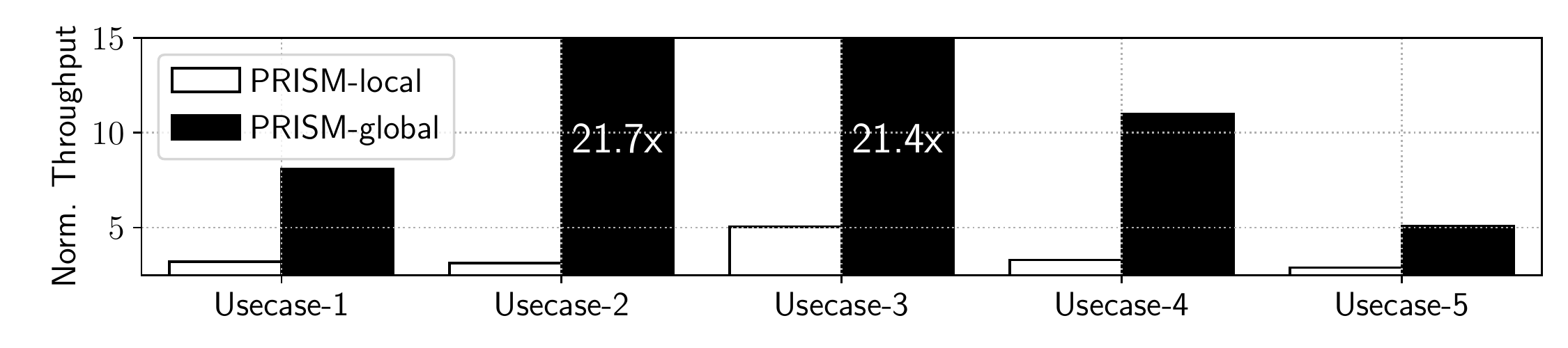}}
	\vspace{-10pt}
	\caption{Throughput normalized to baseline. The first application of a use-case is the current application that is executing on an \nsoc{}. The second application is the one that is invoked by a user at run-time.}
	\vspace{-10pt}
	\label{fig:usecase_results}
\end{figure}

Since no existing schedulers can map use-cases, we created a baseline where the second application of a use-case is mapped by randomly allocating its actors.
Figure~\ref{fig:usecase_results} reports the throughput using \tech{'s} local and global explorations normalized to this baseline for five use-cases.

We observe that \tech{'s} local exploration results in an average 3.5x higher throughput than baseline. This is because during local exploration, \tech{} analyzes all pre-computed schedules of the new application using a probabilistic formulation and selects one that minimizes the resource contention related slowdown for both applications (ongoing and new).
On the other hand, baseline does not incorporate performance when mapping the new application of a use-case. \tech{'s} global exploration results in 3.8x and 13.4x higher throughput than the local exploration and baseline, respectively. The improvement is because during global exploration, \tech{} performs the Hill Climbing heuristic on the merged graph of the two applications to find optimized schedules for both. This results in generating schedules that give higher throughput.

\subsection{Wait Times}
Table~\ref{tab:wait_times} reports wait times for the second application of each use-case when an \nsoc{} is currently executing the first application. Here, the wait time is measured as the time from when a user invokes the second application to the time when the application starts executing on the hardware. This wait time depends on when a schedule is created for the second application. We observe that the local exploration is faster than global exploration (on average, 12x lower wait time). This improves the user experience.

\vspace{-10pt}
\begin{table}[h!]
	\renewcommand{\arraystretch}{0.8}
	\setlength{\tabcolsep}{2pt}
	\caption{Wait times of \tech{'s} local and global explorations.}
	\label{tab:wait_times}
	\vspace{-5pt}
	\centering
	\begin{threeparttable}
	{\fontsize{8}{12}\selectfont
		\begin{tabular}{c|ccccc}
			\hline
			\textbf{\tech{}} & \textbf{Usecase-1} & \textbf{Usecase-2} & \textbf{Usecase-3} & \textbf{Usecase-4} & \textbf{Usecase-5}\\
			\hline
			local & 12s & 21s & 6s & 9s & 14s\\
			global & 223s & 116s & 28s & 138s & 226s\\
			\hline
	\end{tabular}}
	\end{threeparttable}
\end{table}
\vspace{-10pt}

\section{Conclusion}\label{sec:conclusions}
We propose \tech{}, a real-time performance-oriented scheduler for neuromorphic system-on-chips (NSoCs). 
\tech{} operates in four steps.
First, it creates an \ipc{ (IPC)} graph of a machine learning model by considering the mapping of its operations and a self-timed schedule.
Next, it embeds a transaction order for the inter-processor communications into the IPC graph.
Next, it schedules the graph by overlapping communication with the computation.
Finally, it uses a Hill Climbing heuristic to explore the design space of mapping and scheduling IPC graphs to an \nsoc{}, exploiting the platform heterogeneity in improving opportunities for batch, pipeline, and operation parallelism. 
For multi-application uses-cases, \tech{} uses a probabilistic framework to model resource contention related slowdown. It creates a schedule 
to reduce the expected wait time before concurrent operations are scheduled on contending resources.
Our extensive evaluations with 20 machine learning 
workloads and five use-cases show that \tech{} significantly improves the performance per watt for both individual applications 
and 
multi-application use-cases when compared to state-of-the-art schedulers.

\section*{Acknowledgments}
This work is supported by U.S. Department of Energy under Award Number DE-SC0022014 and the National Science Foundation under Awards CCF-1937419 \& CCF-1942697.

\bibliographystyle{IEEEtran}
\IEEEtriggeratref{58} 
\bibliography{commands,disco,external}

\begin{thebibliography}{10}
\providecommand{\url}[1]{#1}
\csname url@samestyle\endcsname
\providecommand{\newblock}{\relax}
\providecommand{\bibinfo}[2]{#2}
\providecommand{\BIBentrySTDinterwordspacing}{\spaceskip=0pt\relax}
\providecommand{\BIBentryALTinterwordstretchfactor}{4}
\providecommand{\BIBentryALTinterwordspacing}{\spaceskip=\fontdimen2\font plus
\BIBentryALTinterwordstretchfactor\fontdimen3\font minus
  \fontdimen4\font\relax}
\providecommand{\BIBforeignlanguage}[2]{{%
\expandafter\ifx\csname l@#1\endcsname\relax
\typeout{** WARNING: IEEEtran.bst: No hyphenation pattern has been}%
\typeout{** loaded for the language `#1'. Using the pattern for}%
\typeout{** the default language instead.}%
\else
\language=\csname l@#1\endcsname
\fi
#2}}
\providecommand{\BIBdecl}{\relax}
\BIBdecl

\bibitem{han2016energy}
B.~Han, A.~Sengupta, and K.~Roy, ``On the energy benefits of spiking deep
  neural networks: A case study,'' in \emph{International Joint Conference on
  Neural Networks (IJCNN)}, 2016.

\bibitem{sengupta2019going}
A.~Sengupta, Y.~Ye, R.~Wang, C.~Liu, and K.~Roy, ``Going deeper in spiking
  neural networks: {VGG} and residual architectures,'' \emph{Frontiers in
  Neuroscience}, vol.~13, p.~95, 2019.

\bibitem{cao2015spiking}
Y.~Cao, Y.~Chen, and D.~Khosla, ``Spiking deep convolutional neural networks
  for energy-efficient object recognition,'' \emph{International Journal of
  Computer Vision}, vol. 113, pp. 54--66, 2015.

\bibitem{maass1997networks}
W.~Maass, ``{Networks of spiking neurons: The third generation of neural
  network models},'' \emph{Neural Networks}, vol.~10, pp. 1659--1671, 1997.

\bibitem{datta2021can}
G.~Datta and P.~A. Beerel, ``Can deep neural networks be converted to ultra
  low-latency spiking neural networks?'' in \emph{Design, Automation, and Test
  in Europe (DATE) Conference and Exhibition}, 2022.

\bibitem{xing2019homeostasis}
F.~Xing, Y.~Yuan, H.~Huo, and T.~Fang, ``Homeostasis-based cnn-to-snn
  conversion of inception and residual architectures,'' in \emph{International
  Conference on Neural Information Processing (ICONIP)}, 2019.

\bibitem{fang2021deep}
W.~Fang, Z.~Yu, Y.~Chen, T.~Huang, T.~Masquelier, and Y.~Tian, ``Deep residual
  learning in spiking neural networks,'' \emph{Conference on Neural Information
  Processing Systems (NeurIPS)}, 2021.

\bibitem{tecs2021}
S.~Song, H.~Chong, A.~Balaji, A.~Das, J.~Shackleford, and N.~Kandasamy,
  ``{DFSynthesizer: Dataflow-based synthesis of spiking neural networks to
  neuromorphic hardware},'' \emph{ACM Transactions on Embedded Computing
  Systems}, vol.~2, pp. 1--4, 2021.

\bibitem{mdpi2022}
A.~Paul, M.~A.~S. Tajin, A.~Das, W.~Mongan, and K.~Dandekar, ``Energy-efficient
  respiratory anomaly detection in premature newborn infants,''
  \emph{Electronics}, pp. 689--694, 2022.

\bibitem{jolpe2018}
A.~Balaji, F.~Corradi, A.~Das, S.~Pande, S.~Schaafsma, and F.~Catthoor,
  ``{Power-accuracy trade-offs for heartbeat classification on neural networks
  hardware},'' \emph{Journal of Low Power Electronics}, vol.~14, pp. 508--519,
  2018.

\bibitem{mead1990neuromorphic}
C.~Mead, ``Neuromorphic electronic systems,'' \emph{Proceedings of the IEEE},
  vol.~78, pp. 1629--1636, 1990.

\bibitem{li2021escalate}
S.~Li, E.~Hanson, X.~Qian, H.~H. Li, and Y.~Chen, ``{ESCALATE: Boosting} the
  efficiency of sparse {CNN} accelerator with kernel decomposition,'' in
  \emph{International Symposium on Microarchitecture (MICRO)}, 2021.

\bibitem{shao2021memory}
Z.~Shao, X.~Chen, L.~Du, L.~Chen, Y.~Du, W.~Zhuang, H.~Wei, C.~Xie, and
  Z.~Wang, ``Memory-efficient {CNN} accelerator based on interlayer feature map
  compression,'' \emph{IEEE Transactions on Circuits and Systems I: Regular
  Papers}, vol.~69, pp. 668--681, 2021.

\bibitem{mrazek2019alwann}
V.~Mrazek, Z.~Vas{\'\i}cek, L.~Sekanina, M.~A. Hanif, and M.~Shafique,
  ``{ALWANN: Automatic layer-wise approximation of deep neural network
  accelerators without retraining},'' in \emph{International Conference on
  Computer-Aided Design (ICCAD)}, 2019.

\bibitem{springer2018}
F.~Catthoor, S.~Mitra, A.~Das, and S.~Schaafsma, ``Very large-scale
  neuromorphic systems for biological signal processing,'' in \emph{CMOS
  Circuits for Biological Sensing and Processing}, 2018.

\bibitem{SB00}
S.~Sriram and S.~Bhattacharyya, \emph{{Embedded} {Multiprocessors};
  {Scheduling} and {Synchronization}}, 2000.

\bibitem{glsvlsi2018}
A.~Das and A.~Kumar, ``Dataflow-based mapping of spiking neural networks on
  neuromorphic hardware,'' in \emph{Great Lakes Symposium on VLSI (GLSVLSI)},
  2018.

\bibitem{isvlsi2019}
A.~Balaji and A.~Das, ``A framework for the analysis of throughput-constraints
  of {SNNs} on neuromorphic hardware,'' in \emph{IEEE Annual Symposium on VLSI
  (ISVLSI)}, 2019.

\bibitem{lctes2020}
S.~Song, A.~Balaji, A.~Das, N.~Kandasamy, and J.~Shackleford, ``{Compiling
  spiking neural networks to neuromorphic hardware},'' in \emph{International
  Conference on Languages, Compilers, and Tools for Embedded Systems (LCTES)},
  2020.

\bibitem{iccad2021}
S.~Song, L.~V. Mirtinti, A.~Das, and N.~Kandasamy, ``A design flow for mapping
  spiking neural networks to many-core neuromorphic hardware,'' in
  \emph{International Conference on Computer-Aided Design (ICCAD)}, 2021.

\bibitem{snpe}
Qualcomm. (2022) {Snapdragon Neural Processing Engine (SNPE)}.

\bibitem{agx}
Nvidia. (2022) {Jetson AGX Xavier}.

\bibitem{ignatov2018ai}
A.~Ignatov, R.~Timofte, W.~Chou, K.~Wang, M.~Wu, T.~Hartley, and L.~Van~Gool,
  ``{AI benchmark: Running deep neural networks on android smartphones},'' in
  \emph{ECCV Workshops}, 2018.

\bibitem{loihi}
M.~Davies, N.~Srinivasa, T.~H. Lin, G.~Chinya, Y.~Cao, S.~H. Choday, G.~Dimou,
  P.~Joshi, N.~Imam, S.~Jain, Y.~Liao, C.~K. Lin, A.~Lines, R.~Liu,
  D.~Mathaikutty, S.~McCoy, A.~Paul, J.~Tse, G.~Venkataramanan, Y.~H. Weng,
  A.~Wild, Y.~Yang, and H.~Wang, ``{Loihi: A neuromorphic manycore processor
  with on-chip learning},'' \emph{IEEE Micro}, vol.~38, pp. 82--99, 2018.

\bibitem{loihi_mapping}
C.-K. Lin, A.~Wild, G.~N. Chinya, T.-H. Lin, M.~Davies, and H.~Wang, ``Mapping
  spiking neural networks onto a manycore neuromorphic architecture,'' in
  \emph{Programming Language Design and Implementation (PLDI)}, 2018.

\bibitem{akida}
BrainChip. (2022) {Akida Neuromorphic System- on-Chip}.

\bibitem{metatf}
BrainChip. (2022) {MetaTF development environment}.

\bibitem{grai}
``{GrAI Chip and GrAIFlow Software},''
  \url{https://www.graimatterlabs.ai/product}, accessed: 2022-05-10.

\bibitem{speck}
Q.~Liu, O.~Richter, C.~Nielsen, S.~Sheik, G.~Indiveri, and N.~Qiao, ``Live
  demonstration: face recognition on an ultra-low power event-driven
  convolutional neural network {ASIC},'' in \emph{CVPR Workshops}, 2019.

\bibitem{spinnaker}
S.~Furber, F.~Galluppi, S.~Temple, and L.~A. Plana, ``The {SpiNNaker}
  project,'' \emph{Proceedings of the IEEE}, vol. 102, pp. 652--665, 2014.

\bibitem{pacman}
F.~Galluppi, S.~Davies, A.~Rast, T.~Sharp, L.~A. Plana, and S.~Furber, ``A
  hierachical configuration system for a massively parallel neural hardware
  platform,'' in \emph{International Conference on Computing Frontiers (CF)},
  2012.

\bibitem{date2022}
M.~L. Varshika, A.~Balaji, F.~Corradi, A.~Das, J.~Stuijt, and F.~Catthoor,
  ``Design of many-core big little $\mu${Brains} for energy-efficient embedded
  neuromorphic computing,'' in \emph{Design, Automation, and Test in Europe
  (DATE) Conference and Exhibition}, 2022.

\bibitem{tianji}
L.~Shi, J.~Pei, N.~Deng, D.~Wang, L.~Deng, Y.~Wang, Y.~Zhang, F.~Chen, M.~Zhao,
  S.~Song \emph{et~al.}, ``Development of a neuromorphic computing system,'' in
  \emph{International Electron Devices Meeting (IEDM)}, 2015.

\bibitem{neutrams}
Y.~Ji, Y.~Zhang, S.~Li, P.~Chi, C.~Jiang, P.~Qu, Y.~Xie, and W.~Chen,
  ``{NEUTRAMS: Neural network transformation and co-design under neuromorphic
  hardware constraints},'' in \emph{International Symposium on
  Microarchitecture (MICRO)}, 2016.

\bibitem{dynapse}
S.~Moradi, N.~Qiao, F.~Stefanini, and G.~Indiveri, ``{A scalable multicore
  architecture with heterogeneous memory structures for dynamic neuromorphic
  asynchronous processors (DYNAPs)},'' \emph{IEEE Transactions on Biomedical
  Circuits and Systems}, vol.~12, pp. 106--122, 2017.

\bibitem{date2018}
A.~Das, Y.~Wu, K.~Huynh, F.~Dell'Anna, F.~Catthoor, and S.~Schaafsma, ``Mapping
  of local and global synapses on spiking neuromorphic hardware,'' in
  \emph{Design, Automation, and Test in Europe (DATE) Conference and
  Exhibition}, 2018.

\bibitem{tvlsi2019}
A.~Balaji, A.~Das, Y.~Wu, K.~Huynh, F.~G. Dell’Anna, G.~Indiveri, J.~L.
  Krichmar, N.~D. Dutt, S.~Schaafsma, and F.~Catthoor, ``Mapping spiking neural
  networks to neuromorphic hardware,'' \emph{IEEE Transactions on Very Large
  Scale Integration (VLSI) Systems}, vol.~28, pp. 76--86, 2019.

\bibitem{esl2020}
A.~Balaji, S.~Song, A.~Das, J.~Krichmar, N.~Dutt, J.~Shackleford, N.~Kandasamy,
  and F.~Catthoor, ``Enabling resource-aware mapping of spiking neural networks
  via spatial decomposition,'' \emph{Embedded Systems Letters}, vol.~13, pp.
  142--145, 2020.

\bibitem{squeezenet}
F.~N. Iandola, S.~Han, M.~W. Moskewicz, K.~Ashraf, W.~J. Dally, and K.~Keutzer,
  ``{SqueezeNet: AlexNet-level accuracy with 50x fewer parameters and< 0.5 MB
  model size},'' \emph{arXiv}, 2016.

\bibitem{inception}
C.~Szegedy, V.~Vanhoucke, S.~Ioffe, J.~Shlens, and Z.~Wojna, ``Rethinking the
  inception architecture for computer vision,'' in \emph{Computer Vision and
  Pattern Recognition Conference (CVPR)}, 2016.

\bibitem{unet}
O.~Ronneberger, P.~Fischer, and T.~Brox, ``{U-net: Convolutional networks for
  biomedical image segmentation},'' in \emph{International Conference on
  Medical Image Computing and Computer-Assisted Intervention (MICCAI)}, 2015.

\bibitem{bert}
J.~Devlin, M.-W. Chang, K.~Lee, and K.~Toutanova, ``{BERT: Pre-training of deep
  bidirectional transformers for language understanding},'' \emph{arXiv
  preprint arXiv:1810.04805}, 2018.

\bibitem{convlstm}
X.~Shi, Z.~Chen, H.~Wang, D.-Y. Yeung, W.-K. Wong, and W.-c. Woo,
  ``{Convolutional LSTM network: A machine learning approach for precipitation
  nowcasting},'' \emph{Conference on Neural Information Processing Systems
  (NeurIPS)}, 2015.

\bibitem{sdfg}
E.~Lee and D.~Messerschmitt, ``{Synchronous data flow},'' \emph{Proceedings of
  the IEEE}, vol.~75, pp. 1235--1245, 1987.

\bibitem{mpsoc}
W.~Wolf, A.~A. Jerraya, and G.~Martin, ``{Multiprocessor system-on-chip (MPSoC)
  technology},'' \emph{IEEE Transactions on Computer-Aided Design of Integrated
  Circuits and Systems}, vol.~27, pp. 1701--1713, 2008.

\bibitem{rosvall2014constraint}
K.~Rosvall and I.~Sander, ``A constraint-based design space exploration
  framework for real-time applications on {MPSoCs},'' in \emph{Design,
  Automation, and Test in Europe (DATE) Conference and Exhibition}.\hskip 1em
  plus 0.5em minus 0.4em\relax IEEE, 2014.

\bibitem{ghamarian2006throughput}
A.~H. Ghamarian, M.~C. Geilen, S.~Stuijk, T.~Basten, B.~D. Theelen, M.~R.
  Mousavi, A.~J. Moonen, and M.~J. Bekooij, ``Throughput analysis of
  synchronous data flow graphs,'' in \emph{International Conference on
  Application of Concurrency to System Design (ACSD)}, 2006.

\bibitem{Stuijk06dac}
S.~Stuijk, M.~Geilen, and T.~Basten, ``Exploring trade-offs in buffer
  requirements and throughput constraints for synchronous dataflow graphs,'' in
  \emph{Design Automation Conference (DAC)}, 2006.

\bibitem{tecs2014}
A.~Das, A.~Kumar, and B.~Veeravalli, ``Energy-aware task mapping and scheduling
  for reliable embedded computing systems,'' \emph{ACM Transactions on Embedded
  Computing Systems}, vol.~13, pp. 1--27, 2014.

\bibitem{dac2013}
A.~K. Singh, A.~Das, and A.~Kumar, ``Energy optimization by exploiting
  execution slacks in streaming applications on multiprocessor systems,'' 2013.

\bibitem{date2014}
A.~Das, A.~Kumar, and B.~Veeravalli, ``{Temperature aware energy-reliability
  trade-offs for mapping of throughput-constrained applications on multimedia
  MPSoCs},'' in \emph{Design, Automation, and Test in Europe (DATE) Conference
  and Exhibition}, 2014.

\bibitem{tpds2015}
A.~Das, A.~Kumar, and B.~Veeravalli, ``Reliability and energy-aware mapping and
  scheduling of multimedia applications on multiprocessor systems,'' \emph{IEEE
  Transactions on Parallel and Distributed Systems}, vol.~27, pp. 869--884,
  2015.

\bibitem{bambha2002intermediate}
N.~Bambha, V.~Kianzad, M.~Khandelia, and S.~S. Bhattacharyya, ``Intermediate
  representations for design automation of multiprocessor {DSP} systems,''
  \emph{Springer Design Automation for Embedded Systems}, vol.~7, pp. 307--323,
  2002.

\bibitem{singh2013mapping}
A.~K. Singh, M.~Shafique, A.~Kumar, and J.~Henkel, ``Mapping on multi/many-core
  systems: survey of current and emerging trends,'' in \emph{Design Automation
  Conference (DAC)}, 2013.

\bibitem{rosenberg1978data}
A.~L. Rosenberg, ``Data encodings and their costs,'' \emph{Acta Informatica},
  1978.

\bibitem{talbi1993hill}
E.-G. Talbi and T.~Muntean, ``Hill-climbing, simulated annealing and genetic
  algorithms: a comparative study and application to the mapping problem,'' in
  \emph{Hawaii International Conference on System Sciences}, 1993.

\bibitem{smitley1988comparative}
D.~L. Smitley and I.~Lee, ``Comparative analysis of hill climbing mapping
  algorithms,'' \emph{Technical Reports (CIS)}, 1988.

\bibitem{icons2021}
A.~Balaji, S.~Song, T.~Titirsha, A.~Das, J.~Krichmar, N.~Dutt, J.~Shackleford,
  N.~Kandasamy, and F.~Catthoor, ``{NeuroXplorer 1.0: An} extensible framework
  for architectural exploration with spiking neural networks,'' in
  \emph{International Conference on Neuromorphic Systems (ICONS)}, 2021.

\bibitem{chen1999segmented}
J.~Chen, W.-B. Jone, J.-S. Wang, H.-I. Lu, and T.-F. Chen, ``Segmented bus
  design for low-power systems,'' \emph{IEEE Transactions on Very Large Scale
  Integration (VLSI) Systems}, vol.~7, pp. 25--29, 1999.

\bibitem{cf2021}
T.~Titirsha, S.~Song, A.~Balaji, and A.~Das, ``On the role of system software
  in energy management of neuromorphic computing,'' in \emph{International
  Conference on Computing Frontiers (CF)}, 2021.

\bibitem{mwscas2020}
S.~Song and A.~Das, ``A case for lifetime reliability-aware neuromorphic
  computing,'' in \emph{IEEE International Midwest Symposium on Circuits and
  Systems (MWSCAS)}, 2020.

\bibitem{tpds2021}
T.~Titirsha, S.~Song, A.~Das, J.~Krichmar, N.~Dutt, N.~Kandasamy, and
  F.~Catthoor, ``Endurance-aware mapping of spiking neural networks to
  neuromorphic hardware,'' \emph{IEEE Transactions on Parallel and Distributed
  Systems}, vol.~33, pp. 288--301, 2021.

\bibitem{dt2022}
A.~Paul, S.~Song, T.~Titirsha, and A.~Das, ``On the mitigation of read
  disturbances in neuromorphic inference hardware,'' \emph{IEEE Design \&
  Test}, 2022.

\bibitem{bojarski2020nvidia}
M.~Bojarski, C.~Chen, J.~Daw, A.~De{\u{g}}irmenci, J.~Deri, B.~Firner,
  B.~Flepp, S.~Gogri, J.~Hong, L.~Jackel \emph{et~al.}, ``The {NVIDIA} pilotnet
  experiments,'' \emph{arXiv preprint arXiv:2010.08776}, 2020.

\end{thebibliography}

\end{document}